%% file: Incremental_Aggregated.tex
\input TEXSHOP_macros_new.tex

%\input TEXSHOP_small_baseline.tex

\def\section#1{\goodbreak\vskip 3pc plus 6pt minus 3pt\leftskip=-2pc
   \global\advance\sectnum by 1\eqnumber=1\subsectnum=0%
\global\examplnumber=1\figrnumber=1\propnumber=1\defnumber=1\lemmanumber=1\assumptionnumber=1 \conditionnumber =1%
   \line{\hfuzz=1pc{\hbox to 3pc{\bf %\the\sectnum.\quad
   \vtop{\hfuzz=1pc\hsize=38pc\hyphenpenalty=10000\noindent\uppercase{\the\sectnum.\quad #1}}\hss}}
			\hfill}
			\leftskip=0pc\nobreak\tenf
			\vskip 1pc plus 4pt minus 2pt\noindent\ignorespaces}
\def\subsection#1{\noindent\leftskip=0pc\tenf
   \goodbreak\vskip 1pc plus 4pt minus 2pt
               \global\advance\subsectnum by 1
   \line{\hfuzz=1pc{\hbox to 3pc{\bf \the\sectnum.\the\subsectnum.
   \vtop{\hfuzz=1pc\hsize=38pc\hyphenpenalty=10000\noindent{\bf #1}}\hss}}
                        \hfill}
   \leftskip=0pc\nobreak\tenf
                        \vskip 1pc plus 4pt minus 2pt\nobreak\noindent\ignorespaces}

%\newcount\conditionnumber
%\def\conditionnum{\global\advance\conditionnumber by1}
%\def\conditionn{\the\sectnum.\the \conditionnumber} 

%%%%%%%%%% REDEFINITION OF BOX SPACING %%%%%%%%%%%%%%%%

\def\texshopbox#1{\boxtext{462pt}{\vskip-1.5pc\pshade{\vskip-1.0pc#1\vskip-2.0pc}}}

%%%%%%%%%%%%%%%%%%%%%%%%%%%%%%%%%%%%%%%%%%%%

\input miniltx

\ifx\pdfoutput\undefined
  \def\Gin@driver{dvips.def} % we are not running PDFTeX
\else
  \def\Gin@driver{pdftex.def} % we are running PDFTeX
\fi

\input graphicx.sty
\resetatcatcode

\long\def\fig#1#2#3{\vbox{\vskip1pc\vskip#1
\prevdepth=12pt \baselineskip=12pt
\vskip1pc
\hbox to\hsize{\hfill\vtop{\hsize=30pc\noindent{\eightbf Figure #2\ }
{\eightpoint#3}}\hfill}}}

\def\show#1{}

\rightheadline{\botmark}

\pageno=1

\rightheadline{\botmark}

\pn {\bf June 2015 (revised September 2015)}\hfill{\bf Report LIDS - 3176}
\bigskip \bigskip\bigskip

\bigskip

\def\longpapertitle#1#2#3{{\bf \centerline{\helbigb
{#1}}}\medskip{\bf \centerline{\helbigb
{#2}}}\bigskip{\bf \centerline{
{#3}}}\bigskip}

\vskip-3pc

\longpapertitle{Incremental Aggregated Proximal and Augmented}{Lagrangian Algorithms}{ {Dimitri P.\ Bertsekas\footnote{\dag}{\ninepoint  Dimitri Bertsekas is with the Dept.\ of Electr.\ Engineering and
Comp.\ Science, and the Laboratory for Information and Decision Systems, M.I.T., Cambridge, Mass., 02139.}}}

\vskip-0.5pc
\centerline{\bf Abstract}
We consider minimization of the sum of a large number of convex functions, and we propose an incremental  aggregated version of the proximal algorithm, which bears similarity to the incremental aggregated gradient and subgradient methods that have received a lot of recent attention. Under cost function differentiability and strong convexity assumptions, we show linear convergence for a sufficiently small constant stepsize. This result also applies to distributed asynchronous variants of the method, involving bounded interprocessor communication delays. 

We then consider dual versions of incremental proximal algorithms, which are incremental augmented Lagrangian methods for separable equality-constrained optimization problems. Contrary to the standard augmented Lagrangian method, these methods admit decomposition in the minimization of the augmented Lagrangian, and update the multipliers far more frequently. Our incremental aggregated augmented Lagrangian methods bear similarity to several known decomposition algorithms, most of which, however, are not incremental in nature: the augmented Lagrangian decomposition algorithm of Stephanopoulos and Westerberg [StW75], and the related methods of Tadjewski [Tad89] and Ruszczynski  [Rus95], and the  alternating direction method of multipliers (ADMM) and more recent variations. We compare these methods in terms of their properties, and highlight their potential advantages and limitations. 

We also address the solution of separable inequality-constrained optimization problems through the use of nonquadratic augmented Lagrangiias such as the exponential, and we dually consider a corresponding incremental  aggregated version of the proximal algorithm that uses nonquadratic regularization, such as an entropy function. We finally propose a closely related linearly convergent method for minimization of large differentiable sums subject to an orthant constraint, which may be viewed as an incremental  aggregated version of the mirror descent method.

\vskip-0.5pc

\xdef\suf{\tl \gr f}

\section{Incremental Gradient, Subgradient, and Proximal Methods}

\vskip-0.5pc

\pn We consider optimization problems with a cost function that consists of additive components:  
$$\eqalign{\hbox{\rm minimize}\quad &
F(x)\; {\buildrel\rm def\over=} \; \sum_{i=1}^mf_i(x)\cr
\hbox{\rm subject to\ \ }
&x\in X,\cr}\eqnum\show{twoo}$$
where $f_i:\rn\mapsto\re$, $i=1,\ldots,m$,  are convex real-valued functions, and $X$ is a closed convex set. We focus on the case where the number of components $m$ is very large, and there is an incentive to use incremental methods that operate on a single component $f_i$ at each iteration, rather than on the entire cost function $F$. Problems of this type arise often in various practical contexts and have received a lot of attention recently. 

Suitable algorithms include the {\it incremental subgradient method} (abbreviated IS), where a  cost component $f_{i_k}$ is selected at iteration $k$, and an arbitrary subgradient $\suf_{i_k}(x_k)$ of $f_{i_k}$ is used in place of a full subgradient of $F$ at $x_k$:\footnote{\dag}
{\ninepoint  Throughout the paper, we will operate within the $n$-dimensional space $\rn$ with the standard Euclidean norm, denoted $\|\cdot\|$. All vectors are considered column vectors and a prime denotes transposition, so $x'x=\|x\|^2$. The scalar coordinates of an optimization vector such as $x$ are denoted by superscripts, $x=(x^1,\ldots,x^n)$, while sequences of iterates are indexed by subscripts. We use $\suf(x)$ to denote a subgradient of a convex function $f$ at a vector $x\in\rn$, i.e, a vector such that $f(z)\ge f(x)+\suf(x)'(z-x)$ for all $z\in\rn$. The choice of $\suf(x)$ from within the set of all subgradients at $x$ will be clear from the context. If $f$ is differentiable at $x$, $\suf(x)$ is the gradient $\gr f(x)$.}
$$x_{k+1}=P_X\big(x_k-\a_k \suf_{i_k}(x_k)\big),\xdef\incrsubgr{\lab}\eqnum\show{twoo}$$
where $\a_k$ is a positive stepsize, and $P_X(\cdot)$ denotes projection on $X$. It is important here that all components are taken up for iteration with equal long-term frequency, using either a cyclic or a random selection scheme. Methods of this type and their properties have been studied for a long time, and the relevant literature, beginning in the 60's, is too voluminous to list here. The author's survey [Ber10] discusses the history of this algorithm, its convergence properties, and its connections with stochastic approximation methods. Generally, a diminishing stepsize $\a_k$ is needed for convergence, even when the components $f_i$ are differentiable. Moreover the convergence rate properties are generally better when the index $i_k$ is selected by randomization over the set $\{1,\ldots,m\}$ than by a deterministic cyclic rule, as first shown by
Nedi\'c and Bertsekas [NeB01]; see also [BNO03].

Another method, introduced by the author in [Ber10] and further studied in [Ber11], [Ber12], is the {\it incremental proximal method\/} (abbreviated IP), 
$$x_{k+1}\in\arg\min_{x\in X}\left\{f_{i_k}(x)+{1\over 2\a_k}\|x-x_k\|^2\right\}.
\xdef\incrprox{\lab}\eqnum\show{twoo}$$
This method relates to the proximal algorithm (Martinet [Mar70], Rockafellar [Roc76a]) in the same way that the IS method \incrsubgr\ relates to the classical nonincremental subgradient method. Similar to the IS method, it is important that all components are taken up for iteration with equal long-term frequency. The theoretical convergence properties of the IS and IP algorithms are similar, but it is generally believed that IP is more robust, a property inherited from its nonincremental counterpart.

It turns out that the structures of the IS and IP methods \incrsubgr\ and \incrprox\ are quite similar. An important fact in this regard is that the IP method \incrprox\ can be equivalently written as 
$$x_{k+1}=P_X\big(x_k-\a_k \suf_{i_k}(x_{k+1})\big),\xdef\incrproxsubgr{\lab}\eqnum\show{twoo}$$
where $\suf_{i_k}(x_{k+1})$ is a {\it special} subgradient of $f_{i_k}$ at the new point $x_{k+1}$ (see Bertsekas [Ber10], Prop.\ 2.1, [Ber11], Prop.\ 1, or [Ber15], Prop.\ 6.4.1). This special subgradient is determined from the  optimality conditions for the proximal maximization \incrprox. For example if $X=\rn$, we have 
$$\suf_{i_k}(x_{k+1})={x_k-x_{k+1}\over \a_k},$$
which is consistent with Eq.\ \incrproxsubgr.
Thus determining the special subgradient $\suf_{i_k}(x_{k+1})$ may be a difficult problem, and in most cases it is preferable to implement the iteration in the proximal form \incrprox\ rather than the projected form \incrproxsubgr. However, the equivalent form of the IP iteration \incrproxsubgr, when compared with the IS iteration \incrsubgr, suggests the close connection between the IS and IP iterations. In fact this connection is the basis for a combination of the two methods to provide flexibility for the case where some of the cost components $f_i$ are well suited for the proximal minimization of Eq.\ \incrprox, while others are not; see  [Ber10], [Ber11], [Ber12].

\subsubsection{Incremental Aggregated Gradient and Subgradient Methods}

\pn Incremental aggregated methods aim to provide a better approximation of a subgradient of the entire cost function $F$, while preserving the economies accrued from computing a single component subgradient at each iteration.
In particular, the aggregated subgradient method (abbreviated IAS), has the form 
$$x_{k+1}=P_X\lf(x_k-\a_k\sum_{i=1}^m \suf_i(x_{\ell_i})\ri),\xdef\distrincraggrproxproj{\lab}\eqnum\show{twoo}$$
where $\suf_{i}(x_{\ell_i})$ is a ``delayed" subgradient of $f_i$ at some earlier iterate $x_{\ell_i}$. We assume that the indexes $\ell_i$ satisfy 
$$k-b\le \ell_i\le k,\qquad \forall\ i,k,\xdef\bounddelays{\lab}\eqnum\show{twoo}$$
where $b$ is a fixed nonnegative integer. Thus the algorithm uses outdated subgradients from previous iterations for the components $f_i$, $i\ne i_k$, and need not compute a subgradient of these components at iteration $k$. 

The IAS method was
first proposed, to our knowledge, by Nedi\'c, Bertsekas, and Borkar [NBB01]. It was motivated primarily by distributed asynchronous solution of dual separable problems, similar to the ones to be discussed in Section 2 (in a distributed asynchronous context, it is natural to assume that subgradients are used with some delays). A convergence result was shown in [NBB01] assuming that the stepsize sequence $\{a_k\}$ is diminishing, and satisfies the standard conditions
$$\sum_{k=0}^\infty \a_k=\infty,\qquad \sum_{k=0}^\infty \a_k^2<\infty.\xdef\dimstepcond{\lab}\eqnum\show{twoo}$$
This result covers the case of iteration \distrincraggrproxproj\ for the case $X=\rn$; the more general case where $X\ne \rn$ admits a similar analysis. We note that distributed algorithms that involve bounded delays in the iterates have a long history, and are common in various distributed asynchronous computation contexts, including gradient-like and coordinate descent methods; see [Ber89], Sections 7.5-7.8.

Note a limitation of this iteration over the IS iteration: one has to store the past subgradients $\suf_{i}(x_{\ell_i})$, $i\ne i_k$. Moreover, whatever effect the use of previously computed subgradients has, it will not be fully manifested until a subgradient of each component has been computed; this is significant when the number of components $m$ is large. We note also that there are other approaches for approximating a full subgradient of the cost function, which aim at computational economies, such as $\e$-subgradient methods (see Nedi\'c and Bertsekas  [NeB10]  and the references quoted there), and surrogate subgradient methods (see Bragin et.\ al [BLY15] and the references quoted there).

The IAS method \distrincraggrproxproj\ contains as a special case the incremental aggregated gradient method (abbreviated IAG) for the case where the components $f_i$ are differentiable:
$$x_{k+1}=x_k-\a_k\sum_{i=1}^m \gr f_i(x_{\ell_i}),\xdef\iagiter{\lab}\eqnum\show{twoo}$$
where $\ell_i\in[k-b,k]$ for all $i$ and $k$. This method has attracted considerable attention
thanks to a particularly interesting convergence result. For the favorable case where the component gradients $\gr f_i$ are Lipschitz continuous and $F$ is strongly convex, it has been shown that  the IAG method  is linearly convergent to the solution with a sufficiently small but constant stepsize $\a_k\equiv \a$. 
This result was first given by  Blatt, Hero, and Gauchman [BHG08], for the case where the cost components $f_i$ are quadratic and the delayed indexes $\ell_i$ satisfy certain restrictions that are consistent with a cyclic selection of components for iteration (see also [AFB06]). The linear convergence result has been subsequently extended for nonquadratic problems and for various forms of the method by several other authors, including Schmidt, Le Roux, and Bach [SLB13], Mairal [Mai13], [Mai14], and Defazio, Caetano, and Domke [DCD14]. Several schemes have been proposed to address the limitation of having to store the past subgradients $\suf_{i}(x_{\ell_i})$, $i\ne i_k$. Moreover, several experimental studies have confirmed the theoretical convergence rate advantage of the IAG method over the corresponding incremental gradient method under the preceding favorable conditions.  The use of arbitrary indexes $\ell_i\in[k-b,k]$ in the IAG method was introduced in the paper by Gurbuzbalaban, Ozdaglar, and Parillo [GOP15], who gave an elegant and particularly simple linear convergence analysis.

\subsubsection{Incremental Aggregated Proximal Algorithm}

\pn In this paper, we consider an incremental aggregated proximal algorithm (abbreviated IAP), which has the form
$$x_{k+1}\in\arg\min_{x\in X}\left\{f_{i_k}(x)+\sum_{i\ne i_k}\suf_{i}(x_{\ell_i})'(x-x_k)+{1\over 2\a_k}\|x-x_k\|^2\right\},
\xdef\incraggrprox{\lab}\eqnum\show{twoo}$$
where $\suf_{i}(x_{\ell_i})$ is a ``delayed" subgradient of $f_i$ at some earlier iterate $x_{\ell_i}$. We assume that the indexes $\ell_i$ satisfy 
the boundedness condition $\ell_i\in[k-b,k]$, cf.\ Eq.\ \bounddelays. Intuitively, the idea is that the term
$$\sum_{i\ne i_k}\suf_{i}(x_{\ell_i})'(x-x_k)$$ 
in the proximal minimization \incraggrprox\ is a linear approximation to the term
$$\sum_{i\ne i_k}f_{i}(x)$$
[minus the constant $\sum_{i\ne i_k}f_i(x_k)$], which would be used in the standard proximal algorithm
$$x_{k+1}\in\arg\min_{x\in X}\left\{F(x)+{1\over 2\a_k}\|x-x_k\|^2\right\}.
\eqnum\show{twoo}$$

It is straightforward to verify the following equivalent form of  the IAP iteration \incraggrprox:
$$x_{k+1}\in\arg\min_{x\in X}\left\{f_{i_k}(x)+{1\over 2\a_k}\|x-z_k\|^2\right\},
\xdef\incraggrproxa{\lab}\eqnum\show{twoo}$$
where
$$z_k=x_k-\a_k\sum_{i\ne i_k}\suf_{i}(x_{\ell_i}).\xdef\incraggrproxb{\lab}\eqnum\show{twoo}$$
In this form the algorithm is executed as a two-step process: first use $x_k$ and preceding subgradients to compute $z_k$ via Eq.\ \incraggrproxb, and then execute an IP iteration starting from $z_k$. Note a limitation of this iteration over the IP iteration, which is shared with other incremental aggregated methods: to keep updating the vector $z_k$, one has to store the past subgradients $\suf_{i}(x_{\ell_i})$, $i\ne i_k$.

Similar to the IP iteration \incrproxsubgr, the IAP iteration \incraggrprox\ and its equivalent form \incraggrproxa-\incraggrproxb\ can be  written as
$$x_{k+1}=P_X\lf(z_k-\a_k\suf_{i_k}(x_{k+1})\ri),\xdef\incraggrproxproj{\lab}\eqnum\show{twoo}$$
so when executing the iteration, we typically can obtain the subgradient $\suf_{i_k}(x_{k+1})$, which can be used in subsequent IAP iterations. For example, in the unconstrained case where $X=\rn$, from Eq.\ \incraggrproxproj, we see that
$$\suf_{i_k}(x_{k+1})={z_k-x_{k+1}\over \a_k}.$$

It is possible to prove various convergence results for the IAP iteration \incraggrprox, or its equivalent forms \incraggrproxa-\incraggrproxb\ and \incraggrproxb-\incraggrproxproj, for the case where the stepsize $\a_k$ is diminishing and satisfies the standard conditions \dimstepcond.
These results are in line with similar results for the IP method, given in  [Ber10], [Ber11], and for the IAS method \distrincraggrproxproj, given in [NBB01]. 
Since the difference between the IAP and  IAS methods is the use of $\suf_{i_k}(x_{k+1})$ in IAP in place of $\suf_{i_k}(x_{\ell_{i_k}})$ in IAS, intuitively, for a diminishing stepsize, the asymptotic performance of the two methods should be similar, and indeed the convergence proofs for the two methods are  fairly similar, under comparable assumptions. We will thus not go into this convergence analysis. 

\subsubsection{Incremental Aggregated Proximal Algorithm for Unconstrained Problems}

\pn In the unconstrained case where $X=\rn$ and the component functions $f_i$ are differentiable, the IAP iteration  \incraggrproxproj\ can be written as 
$$x_{k+1}=x_k-\a_k\lf(\gr f_{i_k}(x_{k+1})+\sum_{i\ne i_k}\gr f_i(x_{\ell_i})\ri).\xdef\incraggrproxfif{\lab}\eqnum\show{twoo}$$
In this case, one may expect similar convergence behavior for the IAP and IAG methods, under favorable conditions which allow the use of a constant stepsize $\a_k\equiv\a$. In particular, we prove the following for the IAP method.

\xdef\propiapd{\propn}\propnum\show{myproposition}

\texshopbox{
\proposition{\propiapd:} Assume that $X=\rn$ and that the functions $f_i$ are convex and differentiable, and satisfy
$$\big\|\gr f_i(x)-\gr f_i(z)\big\|\le L_i\|x-z\|,\qquad \forall\ x,z\in \rn,$$ 
for some constants $L_i$. Assume further that the function $F=\sum_{i=1}^m f_i$ is strongly convex with unique minimum denoted $x^*$. Then there exists $\ol\a>0$ such that for all $\a\in(0,\ol\a]$, the sequence $\{x_k\}$ generated by the IAP iteration 
\incraggrproxfif\ with constant stepsize $\a_k\equiv\a$
converges to $x^*$ linearly, in the sense that $\|x_k-x^*\|\le \g \r^k$ for some scalars $\g>0$ and $\r\in(0,1)$, and all $k$.
}

The proof, given in Section 3, follows closely the one of [GOP15] for the IAG iteration,  and relies on the similarity of the iterations \incraggrproxfif\ and \iagiter\ [the use of the term $\gr f_{i_k}(x_{k+1})$ in place of the term $\gr f_{i_k}(x_{\ell_{i_k}})$]. A key idea is to view the  IAP iteration \incraggrproxfif\ as a gradient method with errors in the calculation of the gradient, i.e.,
$$x_{k+1}=x_k-\a_k\big(\gr F(x_k)+e_k\big),\xdef\graderroeo{\lab}\eqnum\show{twoo}$$
where $\gr F(x_k)=\sum_{i=1}^m\gr f_{i}(x_{k})$, and
$$e_k=\gr f_{i_k}(x_{k+1})-\gr f_{i_k}(x_{k})+\sum_{i\ne i_k}\big(\gr f_{i}(x_{\ell_i})-\gr f_{i}(x_{k})\big),\xdef\graderroet{\lab}\eqnum\show{twoo}$$
and then to appropriately bound the size of the errors $e_k$.  This is similar to known lines of convergence proofs for gradient and subgradient methods with errors.
The proof of Section 3 applies also to a diagonally scaled version of IAP, where a separate but constant stepsize is used for each coordinate. 

\old{
Moreover a more general version of Prop.\ \propiapd\ can be shown for an aggregated proximal iteration that involves errors in the calculation of the delayed gradients. This iteration takes the form
$$x_{k+1}=x_k-\a_k\lf(\gr f_{i_k}(x_{k+1})+\sum_{i\ne i_k}\gr f_i(x_{\ell_i})+w_k\ri),\xdef\erroriter{\lab}\eqnum\show{twoo}$$
where we assume that there exists $\b>0$ such that the following condition
$$\|w_k\|\le \b\sum_{\ell =\max\{0,k-b\}}^{k+1}\|x_\ell-x^*\|,\qquad\forall\ k\ge 1,\xdef\wkcondition{\lab}\eqnum\show{twoo}$$
or the equivalent condition
$$\|w_k\|\le \b\max_{\{0,k-b\}\le \ell\le k+1}\|x_\ell-x^*\|,\qquad\forall\ k\ge 1,$$
is satisfied.
The result of Prop.\ \propiapd\ can then be proved with a very similar proof, which is outlined in Section 3.
As an example, the IAG iteration \iagiter\ can be written in the form \erroriter\ with 
$$w_k=\gr f_{i_k}(x_{\ell_{i_k}})-\gr f_{i_k}(x_{k+1}),$$
and it can be seen that $w_k$ satisfies Eq.\ \wkcondition, since
$$\|w_k\|\le L_{i_k}\|x_{\ell_{i_k}}-x_{k+1}\|\le L_{i_k}\big(\|x_{\ell_{i_k}}-x^*\|+\|x_{k+1}-x^*\|\big).$$
}

We note that the line of proof of Prop.\ \propiapd\ does not readily extend to the constrained case when $X\ne\rn$, nor is it clear whether and under what conditions linear convergence can be proved. In Section 4, however, we will consider an incremental aggregated proximal algorithm that uses a nonquadratic regularization term and seems to cope better with the case of nonnegativity constraints, i.e., $X=\{x\mid x\ge0\}$.

We finally return to the similarity of the IAP method \incraggrprox\ with the IAS method \distrincraggrproxproj, and note that the two methods admit similar distributed asynchronous implementations, which was described in the paper [NBB01]. In this context, we have a central processor executes the proximal iteration \incraggrprox\ for some selected component $f_{i_k}$, while other processors compute subgradients for other components $f_i$ at points $x_{\ell_i}$, which are supplied by the central processor. These subgradients involve a ``delay" that may be unpredictable, hence the asynchronous character of the computation.

\subsubsection{Local Versions of Proximal Algorithms}

\pn While the analysis of this paper requires that $f_i$ and $X$ are convex, there is a straightforward way to extend our incremental proximal methods to nonconvex problems involving twice differentiable functions, which we will describe briefly. The idea is to use a local version of the proximal algorithm, proposed in the author's paper [Ber79] and based on a local version of the Fenchel duality framework given in [Ber78]. The algorithm  applies to the problem
$$\eqalign{&\hbox{minimize\ \ \ }{f(x)}\cr
&\hbox{subject to\ \ }g(x)=0,\cr}\eqnum\show{oneo}
$$
where $f:\rn\mapsto\re$ and $g:\rn\mapsto\re^r$ are twice continuously differentiable functions, such that $f$ is ``locally convex" over the set $\big\{x\mid g(x)=0\big\}$ (this is defined in terms of assumptions that relate to second order sufficiency conditions of nonlinear programming; see [Ber78], [Ber79]). The local proximal algorithm has the form
$$x_{k+1}\in\arg\min_{g(x)=0}\left\{f(x)+{1\over 2\a_k}\|x-x_k\|^2\right\},
\xdef\localincrprox{\lab}\eqnum\show{twoo}$$
where $\a_k$ is sufficiently small to ensure that the function minimized in Eq.\ \localincrprox\ is convex over $\rn$ [not just locally over the set $\big\{x\mid g(x)=0\big\}$]. A Newton-like version of this algorithm was also given in [Ber79].

There is an incremental version of the local proximal iteration \localincrprox\ for
 problems involving sums of functions. In particular, consider the problem 
$$\eqalign{&\hbox{minimize\ \ \ }{\sum_{i=1}^m f_i(x)}\cr
&\hbox{subject to\ \ }g(x)=0,\cr}\eqnum\show{oneo}
$$
where $f_i:\rn\mapsto\re$ and $g:\rn\mapsto\re^r$ are twice continuously differentiable functions, such that each $f_i$ is ``locally convex" over the set $\big\{x\mid g(x)=0\big\}$, for all $i$. This incremental local proximal iteration is 
$$x_{k+1}\in\arg\min_{g(x)=0}\left\{f_{i_k}(x)+{1\over 2\a_k}\|x-x_k\|^2\right\},
\xdef\localincrprox{\lab}\eqnum\show{twoo}$$
where $i_k$ is the index of the cost component that is iterated on. One may also consider an aggregated form of this incremental iteration. The convergence properties of these algorithms are an interesting subject for investigation, which lies, however, outside the scope of the present paper.

There is also another way to combine local proximal and incremental ideas for the case of the (nonconvex) separable problem in the vector $x=(x^1,\ldots,x^m)$, 
$$\eqalign{&\hbox{minimize\ \ \ }{f(x)\; {\buildrel\rm def\over=} \; \sum_{i=1}^m f_i(x^i)}\cr
&\hbox{subject to\ \ }g(x)\; {\buildrel\rm def\over=} \;\sum_{i=1}^m g_i(x^i) =0,\cr}\xdef\seplocal{\lab}\eqnum\show{oneo}$$
where $f_i:\re^{n_i}\mapsto\re$ and $g_i:\re^{n_i}\mapsto\re^r$ are twice continuously differentiable functions, and are such that the problem admits a solution-Lagrange multiplier pair $(x^*,\l^*)$ satisfying standard second order sufficiency conditions. In this approach, also developed in [Ber78], [Ber79], the problem \seplocal\ is converted to the equivalent problem
$$\eqalign{&\hbox{minimize\ \ \ }{\phi_\g(z)\; {\buildrel\rm def\over=} \; \min_{g(x)=0}\lf\{f(x)+{1\over 2\g}\|x-z\|^2\ri\}}\cr
&\hbox{subject to\ \ }z\in\re^{n_1+\cdots+n_m},\cr}\xdef\seplocalconvex{\lab}\eqnum\show{oneo}
$$
where $\g$ is sufficiently small so that for fixed $z$, $f(x)+{1\over 2\g}\|x-z\|^2$ is convex in $x$ locally, for all $x$ in a suitably small neighborhood of $x^*$, i.e., $\g$ should be such that ${1\over \g} I+\gr^2f(x^*)$ is positive definite.
Since the minimization problem \seplocalconvex, which defines $\phi_{\g}(z)$, is separable of the form
$$\eqalign{&\hbox{minimize\ \ \ }{\sum_{i=1}^m \lf(f_i(x^i)+{1\over 2\g}\|x^i-z^i\|^2\ri)}\cr
&\hbox{subject to\ \ }x\in\re^{n_1+\cdots+n_m},\quad \sum_{i=1}^m g_i(x^i) =0,\cr}\xdef\seplocala{\lab}\eqnum\show{oneo}$$
and locally convex in $x$, for fixed $z$ and suitably small values of $\g$, it can be solved using the augmented Lagrangian-based methods of the next section. Denoting $x(z,\g)$ the optimal solution of this problem for  given $z$ and $\g$, it is shown in [Ber79] (Prop.\ 2.1) (see also [Ber78], Prop.\ 2) that $\phi_\g$ is differentiable and
$$\gr \phi_\g(z)={1\over \g}\big(z-x(z,\g)\big).$$
Thus the gradient algorithm 
$$z_{k+1}=z_k-\g\gr \phi_\g(z_k),\xdef\seplocalgrad{\lab}\eqnum\show{oneo}$$
can be written as $z_{k+1}=x(z_k,\g)$ or equivalently, using Eqs.\ \seplocalconvex\ and \seplocala, in the (local) proximal form
$$x_{k+1}\in \arg\min_{\sum_{i=1}^mg_i(x^i)=0}\left\{\sum_{i=1}^m\lf(f_{i}(x^i)+{1\over 2\g}\|x^i-x^i_k\|^2\ri)\right\}.\xdef\seplocalmin{\lab}\eqnum\show{oneo}$$
Note that the above minimization is amenable to decomposition, including solution  using  the  incremental aggregated augmented Lagrangian  and ADMM methods of the next section, assuming $\g$ is sufficiently small to induce the required amount of convexification to make problem \seplocalmin\ convex (locally within a neighborhood of $x^*$).

The convergence properties of this algorithm are developed in [Ber79], based on a local theory of conjugate functions and Fenchel duality developed in [Ber78]. We refer to these papers for a discussion of the local aspects of the minimization \seplocalmin, as well as for the implementation of the Newton iteration 
$$z_{k+1}=z_k-\big(\gr^2 \phi_\g(z_k)\big)^{-1}\gr \phi_\g(z_k),\eqnum\show{oneo}$$
in analogy with the gradient method \seplocalgrad. A further analysis is again outside the scope of the present paper, and is an interesting subject for investigation.

\vskip-1pc

\section{Incremental Augmented Lagrangian Methods}

\pn 
A second objective of this paper is to consider the application of the IP and IAP methods in a dual setting, where they take the form of incremental augmented Lagrangian algorithms for the separable constrained optimization problem
$$\eqalign{&\hbox{minimize\ \ \ }{\sum_{i=1}^mh_i(y^i)}\cr
&\hbox{subject to\ \ }y^i\in Y_i, \ \ i=1,\ldots,m,\ \ \ \ \ \sum_{i=1}^m(A_i y^i-b_i)=0,\cr}\xdef\thrsc{\lab}\eqnum\show{oneo}
$$
as shown in [Ber15], Section 6.4.3.
Here $h_i:\re^{n_i}\mapsto\re$ are convex functions ($n_i$ is a positive integer, which may  depend on $i$), $Y_i$ are nonempty closed convex subsets of $\re^{n_i}$, $A_i$ are given $r\times n_i$ matrices, and $b_i\in\re^r$ are given vectors. The optimization vector is $y=(y^1,\ldots,y^m)$, and our objective is to consider algorithms that allow decomposition in the minimization of the augmented Lagrangian, so that $m$ separate augmented Lagrangian minimizations are performed, each with respect to a single component $y^i$.  Note that the problem \thrsc\ is unaffected by redefinition of the scalars $b_i$, as long as $\sum_{i=1}^m b_i$ is not changed. It may be beneficial to adjust the scalars $b_i$ so that the residuals $A_i y^i-b_i$ are small near the optimal, and this may in fact be attempted in the course of some algorithms as a form of heuristic. 

Following a standard analysis, the dual function for problem \thrsc\ is given by 
$$Q(\l)=\inf_{y^i\in Y_i,\,i=1,\ldots,m}\lf\{\sum_{i=1}^m\big(h_i(y^i)+\l'(A_i y^i-b_i)\big)\ri\},\eqnum\show{oneo}$$
where $\l\in\re^r$ is the dual vector. By decomposing the minimization over the components $y^i$, $Q$ can be expressed in the additive form
$$Q(\l)=\sum_{i=1}^m q_i(\l),$$
where $q_i$ is the concave function
$$q_i(\l)=\inf_{y^i\in Y_i}\big\{h_i(y^i)+\l'(A_i y^i-b_i)\big\},\qquad i=1,\ldots,m.\xdef\lagrangianmin{\lab}\eqnum\show{twoo}$$

\subsubsection{Dual Gradient-Like Methods for Separable Problems}

\pn Assuming that the dual function components $q_i$ are real-valued (which is true for example if $Y_i$ is compact), the dual function $Q(\l)$ can be minimized with the classical subgradient method.\footnote{\dag}{\ninepoint In the case where $q_i$ is not real-valued, the dual function can be maximized over the set $\cap_{i=1}^m\Lambda_i$, where $\Lambda_i=\big\{\l\mid q_i(\l>-\infty\big\}$. This can be done by using incremental constraint projection methods involving  projection or proximal maximization over a single set $\Lambda_i$ at a time. Methods of this type have been proposed in [Ber11], [Ned11], [WaB13], [WaB15], but their discussion is beyond the scope of the present paper.} This method takes the form
$$\l_{k+1}=\l_k+\a_k \sum_{i=1}^m\tl \gr q_i(\l_k),\eqnum\show{twoo}$$
where $\a_k>0$ is the stepsize and the subgradients $\tl\gr q_i(\l_k)$ are obtained as
$$\tl\gr q_i(\l_k)=A_iy_{k+1}^i-b_i,\qquad i=1,\ldots,m,$$
with all components $y^i$ updated according to
$$y_{k+1}^i\in\arg\min_{y^i\in Y_i}\big\{h_i(y^i)+\l_k'(A_i y^i-b_i)\big\},\qquad i=1,\ldots,m.$$

The additive form of the  dual function $Q$ makes it suitable for application of incremental methods, including the IAS method described in Section 1, which in fact was proposed in [NBB01] with the separable problem \thrsc\ in mind. In the case where the components $q_i$ are differentiable [which is true if the infimum in the definition \lagrangianmin\ is attained uniquely for all $\l$], one may also use the IAG method with a constant but sufficiently small stepsize. This is an incremental aggregated version of a classical dual gradient method proposed in the 60s and often attributed to Everett [Eve63]. It takes the form
$$\l_{k+1}=\l_k+\a\lf(\gr q_{i_k}(\l_{k})+\sum_{i\ne i_k}\gr q_i(\l_{\ell_i})\ri);\eqnum\show{twoo}$$
cf.\ Eq.\ \iagiter. The gradient of the dual function component $q_i$ is given by
$$\gr q_i(\l)=A_iy^i(\l)-b_i,$$
where $y^i(\l)$ is the minimizer over $Y_i$ of 
$$f_i(y^i)+\l'A_iy^i,$$
which is assumed to be unique for differentiability of $q_i$. By streamlining the computations using the preceding relations, we see that the iteration has the following form.
 
\texshopbox{{\bf \pn Incremental Aggregated Dual Gradient Iteration (IADG)}
\smskip
\pn Select a component index $i_k$, and update the single component $y^{i_k}$ according to%}\texshopboxnt{\pn 
$$y_{k+1}^{i_k}\in\arg\min_{y^{i_k}\in Y_{i_k}}\big\{h_{i_k}(y^{i_k})+\l_k'A_{i_k}y^{i_k}\big\},\eqnum\show{twoo}$$ 
while keeping the others unchanged, $y^i_{k+1}=y^i_k$ for all $i\ne i_k$. Then update $\l$ according to
$$\l_{k+1}=\l_k+\a\lf(A_{i_k}y_{k+1}^{i_k}+\sum_{i\ne i_k}A_{i}y_{\ell_i}^{i}-b\ri).\eqnum\show{twoo}$$
}

The convergence properties of the method are governed by the known results for the IAG method, which were noted in Section 1.  In particular, we obtain linear convergence with a constant sufficiently small stepsize $\a$, assuming Lipschitz continuity of $\gr q_i$ and strong convexity of $Q$, and that the long-term frequency of updating $y^i$ is the same for all $i$. Note, however, that this linear convergence result cannot be used when the primal problem \thrsc\ has additional convex inequality constraints, because then the corresponding dual problem involves nonnegativity constraints. 

\subsubsection{Augmented Lagrangian-Based Algorithms for Separable Problems}

\pn The nonincremental and incremental subgradient and gradient methods just described are convenient for the purposes of decomposition, but their convergence properties tend to be fragile. On the other hand, the more stable augmented Lagrangian methods have a major drawback: when a quadratic penalty term is added to the Lagrangian function, the resulting augmented Lagrangian
$$\sum_{i=1}^m\big(h_i(y^i)+\l'(A_iy^i-b_i)\big)+{\a_k\over 2}\lf\|\sum_{i=1}^m(A_iy^i-b_i)\ri\|^2$$
is not separable any more, and is not amenable to minimization by decomposition. 
This is a well-known limitation of the augmented Lagrangian approach that has been addressed by a number of authors with various algorithmic proposals, which we will now survey. 

The first proposal of this type was the paper by Stephanopoulos and Westerberg [StW75], which was based on enforced decomposition:  minimizing the augmented Lagrangian separately with respect to each component vector $y^i$, while holding the other components fixed at some estimated values. Minimization over the components $y^i$ is followed by a multiplier update (using the standard augmented Lagrangian formula). The decomposition method of [StW75] attracted considerable attention and motivated further research, including the similarly structured methods by Tadjewski [Tad89] and by Ruszczynski  [Rus95],  which include convergence analyses and give references to earlier works. Our incremental aggregated proximal algorithm bears similarity with the methods of [StW75], [Tad89], and [Rus95]. We note, however, that the methods of [StW75] and [Tad89] were motivated by nonconvex separable problems for which there is a duality gap, while our analysis requires a convex programming structure, where there is no duality gap. The method of [Rus95] is applied to convex separable problems, including linear programming. 

Another method for convex separable problems that uses augmented Lagrangian minimizations is given by Deng, Lai, Peng, and Yin [DLP14], who give several related references, including the paper by Chen and Teboulle [ChT94]. The method is based on the use of primal proximal terms in the augmented Lagrangian (in addition to the quadratic penalty term). This is in the spirit of Rockafellar's proximal method of multipliers [Roc76b], and involves two separate penalty parameters, which for convergence should satisfy certain restrictions. The papers by Hong and Luo [HoL13], and Robinson and Tappenden  [RoT15] also propose algorithms that use primal proximal terms and two penalty parameters, but differ from the algorithm of [DLP14] in that they update the primal variables in Gauss-Seidel rather than Jacobi fashion, while requiring additional assumptions (see also Dang and Lan [DaL15] for a related algorithm). Gauss-Seidel updating is somewhat similar to the incremental mode of iteration of this paper, and based on the results of experiments in [WHM13] and [RoT15], it appears to be beneficial. 

A different possibility to deal with nonconvex separable problems is based on the convexification provided by the local proximal algorithm that was discussed at the end of the preceding section. Its application to nonconvex separable problems is described in [Ber79]; see also Tanikawa and Mukai [TaM85], who proposed a method that aims at improved efficiency relative to the approach of [Ber79]. A discussion of additional proposals of decomposition methods that use augmented Lagrangians is given in the recent paper by Hamdi and Mishra [HaM11].

Still another approach  that has been used to exploit the structure of the separable problem \thrsc\ is the alternating direction method of multipliers (ADMM), a popular method for convex programming, first proposed by Glowinskii and Morocco  [GIM75], and Gabay and Mercier [GaM76], and  further developed by Gabay [Gab79], [Gab83]. This method applies to the problem
$$\eqalign{&\hbox{\hel minimize\  \ \ }f_1(x)+f_2(z)\cr
&\hbox{\hel subject to\ \ }x\in \rn,\,z\in\re^m,\ \ Ax=z,\cr}\eqnum\show{oneo}
$$
where $f_1:\rn\mapsto(-\infty,\infty]$ and $f_2:\re^m\mapsto(-\infty,\infty]$ are closed convex functions, and $A$ is a given $m\times n$ matrix.
The method is better suited than the augmented Lagrangian method for exploiting special structures, including separability, and is capable of decoupling the vectors $x$ and $z$ in the 
augmented Lagrangian 
$$f_1(x)+f_2(z)+\l'(Ax-z)+{\a\over 2}\|Ax-z\|^2.$$
For a discussion of the properties and the many applications of the method, we refer to its extensive literature, including the books [BeT89], Section 3.4.4, [Ber15], Section 5.4, and [BPC11], which give many references. The form of the ADMM for separable problems to overcome the coupling of variables in the augmented Lagrangian minimization was first derived in Bertsekas and Tsitsiklis [BeT89], Section 3.4, pp.\ 249-254 (see also [Ber15], Section 5.4.2). We will describe the form of this specialized ADMM later in this section.

We will now consider the incremental proximal methods IP [cf.\ Eq.\ \incrprox] and IAP [cf.\ Eq.\ \incraggrprox] for maximizing the dual function $\sum_{i=1}^mq_i(\l)$. Taking into account the concavity of the components $q_i$, the IP method takes the form  
$$\l_{k+1}\in\arg\max_{\l\in \re^r}\lf\{q_{i_k}(\l)-{1\over 2\a_k}\|\l-\l_k\|^2\ri\},\xdef\incrproxtt{\lab}\eqnum\show{twoo}$$
where $i_k$ is the index of the component chosen for iteration and $\a_k$ is a positive parameter. This method was given in [Ber15], Section 6.4.3, where it was shown that it can be implemented through the use of decoupled augmented Lagrangian minimizations, each involving a single component vector $y^i$. The
IAP method takes the form  
$$\l_{k+1}\in\arg\max_{\l\in \re^r}\lf\{q_{i_k}(\l)+\sum_{i\ne i_k}\tl \gr q_{i}(\l_{\ell_i})'(\l-\l_k)-{1\over 2\a_k}\|\l-\l_k\|^2\ri\},\xdef\incrproxttagr{\lab}\eqnum\show{twoo}$$
and has not been considered earlier within the dual separable constrained optimization context of this section. The convergence results noted in Section 1 apply to this method. In particular, by Prop.\ \propiapd, the IAP method \incrproxttagr\ is convergent with a sufficiently small constant stepsize, assuming that each $q_i$ is differentiable with Lipschitz continuous gradient and $Q$ is strongly concave. Of course, the differentiability of $q_i$ is a restrictive assumption, and it amounts to attainment of the minimum at a unique point $y^i\in Y_i$ in the definition \lagrangianmin\ of $q_i(\l)$ for all $\l\in\re^r$.

We will now describe how the incremental proximal methods IP and IAP can be implemented in terms of augmented Lagrangian minimizations, which decompose with respect to components $y^i$ and have an incremental character. To this end, we will review the well-known Fenchel duality relation between  proximal and augmented Lagrangian iterations, given first by Rockafellar [Roc73], [Roc76b], and subsequently in many sources, including the author's monograph and textbook accounts [Ber82], Chapter 5, and [Ber15], Section 5.2.1. 

\subsubsection{Duality Between Proximal and Augmented Lagrangian Iterations}

\pn Given a proper convex function $P:\re^r\mapsto(-\infty,\infty]$, let $Q:\re^r\mapsto[-\infty,\infty)$ be the closed proper concave function defined by\footnote{\dag}{\ninepoint Here and later,  for concave functions $Q$, we use 
terminology used for convex functions as applied to $-Q$.}
$$Q(\l)=\inf_{u\in\re^r}\big\{P(u)+\l'u\big\}.\xdef\primalconj{\lab}\eqnum\show{twoo}$$
This is a conjugacy relation, since $Q(\l)=-P^\star(-\l)$, where $P^\star$ is the conjugate convex function of $P$. Moreover, if $P$ is closed, it  can be recovered from $Q$ using the conjugacy theorem,
$$P(u)=P^{\star\star}(u)=\sup_{\l\in\re^r}\big\{\l'u+Q(-\l)\big\},\xdef\dualconj{\lab}\eqnum\show{twoo}$$
where $P^{\star\star}$ is the conjugate convex function of $P^\star$ (see, e.g., [Ber09], Prop.\ 1.6.1).

A key fact, assuming that $P$ is closed, is that the proximal iteration 
$$\l_{k+1}\in\arg\max_{\l\in \re^r}\lf\{Q(\l)-{1\over 2\a_k}\|\l-\l_k\|^2\ri\},\xdef\dualprox{\lab}\eqnum\show{twoo}$$
can be equivalently implemented in two steps as
$$u_{k+1}\in\arg\min_{u\in\re^r}\lf\{P(u)+\l_k'u+{\a_k\over 2}\|u\|^2\ri\},\xdef\primalprox{\lab}\eqnum\show{twoo}$$
followed by
$$\l_{k+1}=\l_k+\a_k u_{k+1};\xdef\dualiteru{\lab}\eqnum\show{twoo}$$
see, e.g., [Ber15], Section 5.2.1. Moreover, $u_{k+1}$ is a subgradient of $Q$ at $\l_{k+1}$:
$$u_{k+1}=\tl \gr Q(\l_{k+1}).\xdef\uasgrad{\lab}\eqnum\show{twoo}$$
These relations are shown by straightforward application of the Fenchel duality theorem to the maximization of Eq.\ \dualprox, which involves the sum of the concave functions $Q$ and $-(1/2\a_k)\|\l-\l_k\|^2$. The closedness of $P$ is used both to ensure that the duality relation \dualconj\ holds, and to guarantee that the minimum in Eq.\ \primalprox\ is attained. Note that Eq.\ \primalprox\ has the form of an augmented Lagrangian minimization relating to the (somewhat contrived) problem of minimizing $P$ subject to the equality constraint $u=0$. 

\subsubsection{Augmented Lagrangian Method}

\pn We will now translate the duality between the proximal and augmented Lagrangian iterations just described to the  
constrained optimization context, setting the stage for using this duality in an incremental context. Consider a generic convex programming problem of the form
$$\eqalign{&\hbox{minimize\ \ \ }H(y)\cr
&\hbox{subject to\ \ }y\in Y,\ \ \ \ \ Ay-b=0,\cr}\xdef\genconstrproblem{\lab}\eqnum\show{oneo}
$$
where $H:\rn\mapsto\re$ is a convex function, $Y$ is a convex set, $A$ is an $r\times n$ matrix, and $b\in \re^r$.
Consider also the corresponding primal and dual functions
$$P(u)=\inf_{y\in Y,\,Ay-b=u}H(y),\qquad Q(\l)=\inf_{y\in Y}\big\{H(y)+\l'(Ay-b)\big\},$$
which are convex and concave, respectively. 
We assume that $P$ is closed and proper, and that the optimal value of the problem is finite, so that $Q$ is also closed proper and concave, and there is no duality gap (see [Ber09], Section 4.2). 

There is a well-known relation between the primal and dual functions. In particular, $Q$ has the equivalent form
$$Q(\l)=\inf_{u\in\re^r}\inf_{y\in Y,\, Ay-b=u}\big\{H(y)+\l'(Ay-b)\big\}=\inf_{u\in \re^r}\big\{P(u)+\l'u\big\},$$
so $P$ and $Q$ satisfy the conjugacy relation \primalconj. Based on the preceding discussion [cf.\ \primalconj-\uasgrad], it follows that the proximal iteration \dualprox\ can be equivalently written as the two-step process \primalprox-\dualiteru
$$u_{k+1}\in\arg\min_{u\in\re^r}\lf\{P(u)+\l_k'u+{\a_k\over 2}\|u\|^2\ri\},\xdef\primalprox{\lab}\eqnum\show{twoo}$$
followed by
$$\l_{k+1}=\l_k+\a_k u_{k+1}.\xdef\dualmultiter{\lab}\eqnum\show{twoo}$$
Moreover, from Eqs.\ \dualiteru\ and \uasgrad, we have
$$u_{k+1}={\l_{k+1}-\l_k\over \a_k}=\tl \gr Q(\l_{k+1}).\xdef\uitersubgrad{\lab}\eqnum\show{twoo}$$

We will now write the iteration \primalprox-\dualmultiter\ in terms of the augmented Lagrangian, and obtain the classical (first order) augmented Lagrangian method.
Using the definition of the primal function $P$, we see that the minimization in Eq.\ \primalprox\ can be written as
$$\eqalign{\inf_{u\in\re^r}&\lf\{P(u)+{\l_k}' u+{\a_k\over 2}\|u\|^2\ri\}\cr
&=\inf_{u\in\re^r}\lf\{\inf_{y\in Y,\,Ay-b=u}\bl\{H(y)\br\}+{\l_k}' u+{\a_k\over 2}\|u\|^2\ri\}\cr
&=\inf_{u\in\re^r}\inf_{y\in Y,\,Ay-b=u}\lf\{H(y)+\l_k'(Ay-b)+{\a_k\over 2}\|Ay-b\|^2\ri\}\cr
&=\inf_{y\in Y}\lf\{H(y)+\l_k'(Ay-b)+{\a_k\over 2}\|Ay-b\|^2\ri\}\cr
&=\inf_{y\in Y}L_{\a_k}(y,\l_k),\cr}$$
where for any $\a>0$, $L_\a$ is the augmented Lagrangian function
$$L_\a(y,\l)=H(y)+\l'(Ay-b)+{\a\over 2}\|Ay-b\|^2,\qquad y\in\rn,\ \l\in\re^r.\eqnum\show{twoo}$$
From the preceding calculation it also follows that  for any $y_{k+1}\in Y$ that minimizes the augmented Lagrangian  over $Y$:
$$y_{k+1}\in\arg\min_{y\in Y} L_{\a_k}(y,\l_k),\xdef\xminauglagr{\lab}\eqnum\show{twoo}$$
we have $u_{k+1}=Ay_{k+1}-b$, and the iteration \dualmultiter\ can be equivalently written as the multiplier iteration
$$\l_{k+1}=\l_k+\a_k (Ay_{k+1}-b).\xdef\multiter{\lab}\eqnum\show{twoo}$$
This is precisely the (first order) augmented Lagrangian method. It is equivalent to the proximal iteration 
$$\l_{k+1}\in\arg\max_{\l\in \re^r}\lf\{Q(\l)-{1\over 2\a_k}\|\l-\l_k\|^2\ri\},$$
[cf.\ Eq.\ \dualprox]. In view of Eqs.\ \uitersubgrad\ and \multiter, it can also be written in the gradient-like form
$$\l_{k+1}=\l_k+\a_k \tl \gr Q(\l_{k+1}),\xdef\multiter{\lab}\eqnum\show{twoo}$$
where $\tl\gr Q(\l_{k+1})$, the special subgradient of $Q$ at $\l_{k+1}$, is given by
$$ \tl \gr Q(\l_{k+1})=Ay_{k+1}-b.\xdef\multitert{\lab}\eqnum\show{twoo}$$

Note that the minimizing $y_{k+1}$ in Eq.\ \xminauglagr\ need not exist or be unique. Its existence must be assumed in some way, e.g., by assuming that $H$ has compact level sets. As an example, it can be verified that for the two-dimensional/single constraint problem of minimizing  $H(y)=e^{y^1}$, subject to $y^1+y^2=0$, $y^1\in\re$, $y^2\ge0$, the dual optimal solution is $\l^*=0$, but there is no primal optimal solution. For this problem, the augmented Lagrangian algorithm will generate sequences $\{\l_k\}$ and $\{y_k\}$ such that $\l_k\to0$ and $y_k\to-\infty$.

\subsubsection{Incremental Augmented Lagrangian Methods}

\pn The duality between the proximal and augmented Lagrangian minimizations outlined above is generic, and holds in other related contexts, based on a similar use of the Fenchel duality theorem. In the context of the separable problem \thrsc, it holds in an incremental form where $Q(\l)$ is replaced by 
$$q_{i_k}(\l),$$
as in the IP iteration  \incrproxtt, or is replaced by 
$$q_{i_k}(\l)+\sum_{i\ne i_k}\tl \gr q_{i}(\l_{\ell_i})'(\l-\l_k),$$
as in the IAP iteration \incrproxttagr. We refer to these two methods as the {\it incremental augmented Lagrangian method\/} (abbreviated IAL), and  the {\it incremental aggregated augmented Lagrangian method\/} (abbreviated IAAL).

Based on the discussion of the algorithm \xminauglagr-\multiter, the IAL method,
$$\l_{k+1}\in\arg\max_{\l\in\re^r}\lf\{q_{i_k}(\l)-{1\over 2\a_k}\|\l-\l_k\|^2\ri\},$$
 can be implemented as follows, as already noted in [Ber15], Section 6.4.3.
 
\texshopbox{{\bf \pn Incremental Augmented Lagrangian Iteration (IAL)}
\smskip
\pn Select a component index $i_k$, and update the single component $y^{i_k}$ according to
$$y_{k+1}^{i_k}\in\arg\min_{y^{i_k}\in Y_{i_k}}\lf\{h_{i_k}(y^{i_k})+\l_k'(A_{i_k}y^{i_k}-b_{i_k})+{\a_k\over 2}\|A_{i_k}y^{i_k}-b_{i_k}\|^2\ri\},\xdef\incrxminauglagr{\lab}\eqnum\show{twoo}$$
while keeping the others unchanged, $y^i_{k+1}=y^i_k$ for all $i\ne i_k$. Then update $\l$ according to 
$$\l_{k+1}=\l_k+\a_k (A_{i_k}y_{k+1}^{i_k}-b_{i_k}).\xdef\incrmultiter{\lab}\eqnum\show{twoo}$$
}

As in the IP method, all component indexes should be selected for iteration in Eq.\ \incrxminauglagr\ with equal long-term frequency. Note that the augmented Lagrangian minimization is decoupled with respect to the components $y^i$, thus overcoming the major limitation of the augmented Lagrangian approach for separable problems.

To derive the IAAL method, we use the equivalent form \incraggrproxa-\incraggrproxb\ of the IAP algorithm. We see then that the method has similar form to the  IAL method, except that $\l_k$ is first translated by a multiple of the sum of the delayed subgradients. In particular, the IAAL iteration takes the form
$$\l_{k+1}\in\arg\max_{\l\in\re^r}\lf\{q_{i_k}(\l)- {1\over 2\a_k}\|\l-\nu_k\|^2\right\},$$
where
$$\nu_k=\l_k+\a_k \sum_{i\ne i_k}\tl \gr q_i(\l_{\ell_i}).\xdef\nukequation{\lab}\eqnum\show{twoo}$$
Applying the relations \xminauglagr-\multiter, it follows that we can write the IAAL iteration in two steps:
Select a component index $i_k$, and update the single component $y^{i_k}$ according to
$$y_{k+1}^{i_k}\in\arg\min_{y^{i_k}\in Y_{i_k}}\lf\{h_{i_k}(y^{i_k})+\nu_k'(A_{i_k}y^{i_k}-b_{i_k})+{\a_k\over 2}\|A_{i_k}y^{i_k}-b_{i_k}\|^2\ri\},\xdef\agrauglagrmin{\lab}\eqnum\show{twoo}$$
while keeping the others unchanged, $y^i_{k+1}=y^i_k$ for all $i\ne i_k$. Then update $\l$ according to 
$$\l_{k+1}=\nu_k+\a_k(A_{i_k}y_{k+1}^{i_k}-b_{i_k}).\xdef\agrmultiter{\lab}\eqnum\show{twoo}$$

Note that the subgradients $\tl \gr q_i(\l_{\ell_i})$,  needed for the computation of $\nu_k$ in Eq.\ \nukequation, are generated by
$$\tl \gr q_i(\l_{\ell_i})=A_iy^i_{\ell_i}-b_i,\qquad \forall\ i\ne i_k,$$
[cf.\ Eq.\ \multitert]. Thus by streamlining the preceding relations, we see that the IAAL updates are written as 
$$y_{k+1}^{i_k}\in\arg\min_{y^{i_k}\in Y_{i_k}}\lf\{h_{i_k}(y^{i_k})+\l_k'(A_{i_k}y^{i_k}-b_{i_k})+{\a_k\over 2}\lf\|A_{i_k}y^{i_k}-b_{i_k}+\sum_{i\ne i_k}(A_{i}y_{\ell_i}^{i}-b_{i})\ri\|^2\ri\},$$
$$\l_{k+1}=\l_k+\a_k\lf(A_{i_k}y_{k+1}^{i_k}-b_{i_k}+\sum_{i\ne i_k}(A_{i}y_{\ell_i}^{i}-b_{i})\ri).$$
If we denote
$b=\sum_{i=1}^m b_i,$
and neglect the constant term $-\l_k'b_{i_k}$ from the augmented Lagrangian, we can write the iteration in a way that it depends on the scalars $b_i$ only through their sum $b$.

\texshopbox{{\bf \pn Incremental Aggregated Augmented Lagrangian (IAAL) Iteration}
\smskip
\pn Select a component index $i_k$, and update the single component $y^{i_k}$ according to
$$y_{k+1}^{i_k}\in\arg\min_{y^{i_k}\in Y_{i_k}}\lf\{h_{i_k}(y^{i_k})+\l_k'A_{i_k}y^{i_k}+{\a_k\over 2}\lf\|A_{i_k}y^{i_k}+\sum_{i\ne i_k}A_{i}y_{\ell_i}^{i}-b\ri\|^2\ri\},\xdef\agrauglagrmint{\lab}\eqnum\show{twoo}$$
while keeping the others unchanged, $y^i_{k+1}=y^i_k$ for all $i\ne i_k$. Then update $\l$ according to 
$$\l_{k+1}=\l_k+\a_k\lf(A_{i_k}y_{k+1}^{i_k}+\sum_{i\ne i_k}A_{i}y_{\ell_i}^{i}-b\ri).\xdef\agrmultitert{\lab}\eqnum\show{twoo}$$
}

By comparing the IAL method \incrxminauglagr-\incrmultiter\ with the IAAL method \agrauglagrmint-\agrmultitert, we see that they require comparable computations per iteration. While the IAL method requires a diminishing stepsize $\a_k$ for convergence, the  IAAL method can converge with a constant stepsize, assuming that the dual function components have Lipschitz continuous gradients, and the dual function is strongly concave (cf.\ Prop.\ \propiapd). Intuitively, if it can use a constant stepsize, the IAAL method should be asymptotically more effective than the IAL method. Of course, if $Q$ is not strongly convex (as for example in the important case where $Q$ is polyhedral, which arises in integer programming), our analysis guarantees the convergence of the  IAAL method only if the stepsize $\a_k$ is diminishing. In this case it is unclear which of the IAL and IAAL methods is more effective on a given problem.
 
Both the IAL and IAAL algorithms require an initial multiplier $\l_0$. Regarding the delayed indexes $\ell_i$ in the IAAL algorithm, if the iteration is executed at a single processor, it is most appropriate to choose
$\ell_i$ to be the iteration index at which the component $y^i$ was last changed prior to the current index $k$, so $\ell_i\le k$  (if a component $y^i$ has not yet been updated prior to $k$, we take $\ell_i=0$ and let $y_0^i$ be some initial choice for $y^i$). In this case, the formal statement of the  IAAL method is again given by Eqs.\ \agrauglagrmin-\agrmultiter, with $\ell_i$ replaced by $k$ for all $i\ne i_k$. However, a different value of $\ell_i$ may apply if the iteration is executed in a  distributed asynchronous computing environment, as in the corresponding IAS method of [NBB01].

Note that the multiplier $\l_k$ is updated each time a component $y^i$ is updated, which suggests that the stepsize $\a_k$ should be chosen carefully, possibly through some experimentation. Moreover, the strong convexity assumption of $Q$ is essential for the convergence of the method with a constant stepsize. Indeed a three-dimensional example by Chen, He, Ye, and Yuan [CHH14] can be used to show that the IAAL algorithm need not converge for any value of constant stepsize if the strong convexity assumption is violated.\footnote{\dag}{\ninepoint  While the paper [CHH14] is entitled ``The Direct Extension of ADMM  for Multi-Block Convex Minimization Problems ...," it considers an algorithm that is not a special case of ADMM, so a convergence counterexample is possible. A correct specialization of ADMM for separable problems (dating from 1989 but unknown to the authors of [CHH14]) will be given shortly, and is convergent under the same broadly applicable conditions as ADMM.} An alternative possibility is to perform a batch of component updates $y^i$ of the form \agrauglagrmin\ between multiplier updates of the form \agrmultiter. For example, one may restructure the IAAL iteration so that it consists of a full cycle of updates of $y^1,\ldots,y^m$, sequentially according to Eq.\ \agrauglagrmint, to obtain $y^i_{k+1}$, $i=1,\ldots,m$, and only then to update $\l$ according to
$$\l_{k+1}=\l_k+\a_k\lf(\sum_{i=1}^mA_{i}y_{k+1}^{i}-b\ri).$$
Note that this sequential update of $y^1,\ldots,y^m$  according to Eq.\ \agrauglagrmint\ amounts to a cycle of coordinate descent iterations for minimizing the augmented Lagrangian. Therefore, this variant of the IAAL iteration may be viewed as an implementation of the augmented Lagrangian method with approximate minimization of the augmented Lagrangian using coordinate descent.
An algorithm of this type may be interesting and has been suggested in the past (see Bertsekas and Tsitsiklis [BeT89], Example 4.4, and Eckstein [Eck12]). Its linear convergence has been shown under certain assumptions by Hong and Luo [HoL13]. The algorithm is worthy of further investigation, particularly in view of favorable computational results given by Wang, Hong, Ma, and Luo [WHM13]. Let us also note that the work by Hong, Chang, Wang, Razaviyayn, Ma, and Luo  [HCW14] derives an algorithm for the separable problem \thrsc\ that is quite similar to the IAAL algorithm, using different assumptions and line of development. The paper [HCW14] proves convergence but not a linear convergence rate result.

\vskip-0.5pc

\subsubsection{Comparison with ADMM}

\pn We will now compare the IAAL iteration with  the ADMM.
We note that there is a well-known connection of the ADMM and augmented Lagrangian methods, which was clarified long ago through a series of papers. In particular, Lions and Mercier [LiM79] proposed a splitting algorithm for finding a zero of the sum of two maximal monotone operators, known as the Douglas-Ratchford algorithm.  It turns out that this algorithm contains as a special case the ADMM, as shown in [Gab83]. The paper by Eckstein and Bertsekas [EcB92] showed that the general form of the proximal algorithm for finding a zero of maximal monotone operator, proposed by Rockafellar [Roc76a],  [Roc76b], contains as a special case the Douglas-Ratchford algorithm and hence also the ADMM. Thus the ADMM and the augmented Lagrangian method have a common ancestry: they are both special cases of   the general form of the proximal algorithm for finding a zero of a maximal monotone operator. The common underlying structure of the two methods is reflected in similar formulas, but ADMM has the advantage of flexibility to allow decomposition, at the expense of a typically slower practical convergence rate.

A convenient decomposition-based form of ADMM for the separable problem \thrsc\ was derived (together with the corresponding coordinate descent version of the augmented Lagrangian method) in [BeT89], Section 3.4 and Example 4.4 (see also [Ber15], Section 5.4.2). Wang, Hong, Ma, and Luo [WHM13], apparently unaware of this form of ADMM, give related algorithms (referred to as Algorithms 2 and 3 in their paper), which, however, involve updating $m$ multiplier vectors in place of the single multiplier update of the following algorithm. At iteration $k$, and given $\l_k$, the ADMM algorithm of [BeT89]  generates $\l_{k+1}$ as follows.

\texshopbox{{\bf \pn ADMM Iteration for Separable Problems}
\smskip
\pn Perform a separate augmented Lagrangian minimization over $y^i$, for each $i=1,\ldots,m$,
$$y_{k+1}^{i}\in\arg\min_{y^{i}\in Y_{i}}\lf\{h_{i}(y^{i})+\l_k'A_{i}y^{i}+{\a\over 2}\lf\|A_{i}y^{i}-A_iy_k^i+{1\over m}\lf(\sum_{j=1}^m A_jy_k^j-b\ri)\ri\|^2\ri\},\qquad i=1,\ldots,m,\xdef\admmauglagrmin{\lab}\eqnum\show{twoo}$$
and  then update $\l_k$ according to
$$\l_{k+1}=\l_k+{\a\over m}\lf(\sum_{i=1}^m A_iy_{k+1}^i-b\ri).\xdef\admmmultiter{\lab}\eqnum\show{twoo}$$
}

Note that contrary to the augmented Lagrangian method, where the best strategy for adjusting $\a$ is usually clear, see e.g., [Ber82], there is no clear way to adjust the parameter $\a$ to improve performance in ADMM. As a result for efficiency $\a$ is often determined by  trial and error.
A closely related but more refined form of ADMM, also derived in [BeT89a], Section 3.4, Example 4.4, aims to improve the parameter selection by exploiting the structure of the matrices $A_i$. It uses a coordinate-dependent parameter ${\a\over m_j}$ in  iteration \admmmultiter, in place of $\a/m$, where $m_j$ is the number of submatrices $A_i$ that have nonzero $j$th row. In this version, the multiplier update essentially involves diagonal scaling. The iteration maintains  additional vectors $z_k^i\in\re^r$, $i=1,\ldots,m$, which represent estimates of $A_iy^i$ at the optimum, and has the following form, where $A_{ji}$ denotes the $j$th row of the matrix $A_i$.

\texshopbox{{\bf \pn Diagonally Scaled ADMM Iteration for Separable Problems}
\smskip
\pn Perform a separate augmented Lagrangian minimization over $y^i$, for each $i=1,\ldots,m$,%}\texshopboxnt{\pn
$$y_{k+1}^{i}\in\arg\min_{y^{i}\in Y_{i}}\lf\{h_{i}(y^{i})+\l_k'A_{i}y^{i}+{\a\over 2}\lf\|A_{i}y^{i}-z_k^i\ri\|^2\ri\},\qquad i=1,\ldots,m,\xdef\admmauglagrminsc{\lab}\eqnum\show{twoo}$$%}\texshopboxnt{\pn
and  then update $\l_k$ and $z_k$ according to
$$\l_{k+1}^j=\l_k^j+{\a\over m_j}\lf(\sum_{i=1}^m A_{ji}y_{k+1}^i-b_j\ri),\qquad j=1,\ldots,r,\eqnum\show{twoo}$$
$$z_{k+1}^i=A_iy_{k+1}^i+{\l_k-\l_{k+1}\over \a},\qquad i=1,\ldots,m.\xdef\admmzitersc{\lab}\eqnum\show{twoo}$$
}
 
Note that the preceding two ADMM iterations coincide when there is no nonzero row in any of the matrices $A_i$, i.e., $m_j=m$ for all $j$. In comparing the IAAL iteration \agrauglagrmint-\agrmultitert, and the ADMM iterations \admmauglagrmin-\admmmultiter\ and \admmauglagrminsc-\admmzitersc, we note that they involve fairly similar operations. In particular, the ADMM mutiplier update \admmmultiter\  approximates an average (over a full cycle of $m$ components) of the IAAL multiplier updates \agrmultitert, and is executed $m$ times less frequently; this is reminiscent of the difference between the proximal and incremental proximal iterations. The different multiplier update frequencies of IAAL and ADMM suggests that assuming IAAL converges, its stepsize $\a_k$ should be chosen much smaller than the stepsize $\a$ in ADMM,
say 
$$\a_k\in\lf[{\a\over m},{\a\over m^2}\ri],$$
as a crude approximation, for comparable performance. There are also two other major differences:
\nitem{(a)} The ADMM iterations have guaranteed convergence for any constant stepsize $\a$, and under weaker conditions (differentiability of $q_i$ and  strong convexity of $Q$ are not required). On the other hand the IAAL method requires a diminishing stepsize in general, or (under Lipschitz continuity of $\gr q_i$ and strong convexity of $Q$) a constant stepsize that is not arbitrary, but must be sufficiently small. 
\nitem{(b)} In the IAAL method a {\it single} component $y^i$ is updated at each iteration, while in the ADMM {\it all} components $y^i$ are updated. For some problems, this may work in favor of  IAAL, particularly for large $m$, a case that generally seems to favor incremental methods. 

\smskip
\pn Thus for the separable problems of this section, one may roughly view the IAAL method as an incremental variant of ADMM, where the advantage of incrementalism may be offset by less solid convergence properties. A computational comparison of the two methods will be helpful in clarifying their relative merits.

The diagonally scaled ADMM iteration \admmauglagrminsc-\admmzitersc\ suggests also a similar diagonal scaling for the IAAL iteration. The simplest way to accomplish this is to use the  IAAL method \agrauglagrmint-\agrmultitert\ after scaling the constraints, i.e., after multiplying the $r$ constraint equations with different scaling factors, which in turn will introduce diagonal scaling for the dual variables. Proposition \propiapd\ will still apply under this form of scaling, assuming Lipschitz continuity of $\gr q_i$ and strong convexity of $Q$.

\subsubsection{Comparison with the Methods of Tadjewski [Tad89] and Ruszczynski  [Rus95]}

\pn The methods of [Tad89] and [Rus95] are motivated by the earlier algorithm of [StW75], and apply to the separable constrained optimization problem of this section. They are  similar to each other, but use different assumptions. The method of [Tad89] requires differentiability and second order sufficiency assumptions, but applies to nonconvex separable problems that may have a duality gap, while the method of [Rus95] applies to  separable problems with convex, possibly nondifferentiable cost function. These methods are also similar to our IAAL method \agrauglagrmin-\agrmultiter, but they use different approximations of the quadratic penalty terms. In particular, instead of the vectors $y_{\ell_i}^{i}$ that appear in Eqs.\ \agrauglagrmin\ and \agrmultiter, they use other terms that are iteratively adjusted, with the aim to improve the approximation of the quadratic penalty terms of  the standard augmented Lagrangian. Both papers [Tad89] and [Rus95] provide a convergence analysis, involving suitable choices of various parameters, although the convergence results obtained are not as strong as the ones for ADMM. A major difference of the methods of [Tad89] and [Rus95] from our IAAL method is that, like  the ADMM, they update all the components $y^i$ simultaneously at each iteration, so they are not incremental in character.

\vskip-1pc

\section{Proof of Proposition \propiapd}

\pn Similar to other convergence proofs of incremental gradient methods, including the one of [GOP15] for the IAG method that we follow, the proof of Prop.\ \propiapd\ is based on viewing the IAP iteration  with constant stepsize $\a_k\equiv\a$,
$$x_{k+1}=x_k-\a\lf(\gr f_{i_k}(x_{k+1})+\sum_{i\ne i_k}\gr f_i(x_{\ell_i})\ri),\xdef\incraggrproxt{\lab}\eqnum\show{twoo}$$
as a gradient method with errors in the calculation of the gradient [cf.\ Eqs.\ \graderroeo, \graderroet]. To deal with the delays in the iterates, we use the following lemma, due to Feyzmahdavian,  Aytekin, and Johansson [FAJ14] (which is also used in the convergence proof of [GOP15]):
\xdef\lemmaone{\lemman}\lemmanum\show{mylemma}

\texshopbox{\lemma{\lemmaone:}
Let $\{\b_k\}$ be a nonnegative sequence satisfying%}\texshopboxnt{\pn 
$$\b_{k+1}\le p\b_k+q\max_{\max\{0,k-d\}\le \ell\le k}\b_\ell,\qquad \forall\ k=0,1,\ldots,$$%}\texshopboxnt{\pn 
for some positive integer $d$ and nonnegative scalars $p$ and $q$ such that $p + q < 1$. Then we have
$$\b_k\le \r^k\b_0,\qquad \forall\ k=0,1,\ldots,$$
where $\r=(p+q)^{1\over 1+d}$.
}

In the following proof we take the stepsize $\a$ as small as is needed for the various calculations to be valid. Also for convenience in expressing various formulas involving delays, we consider the algorithm for large enough iteration indexes, so that all the delayed iteration indexes in the following calculations are larger than 0 (for this it will be sufficient to consider the algorithm as starting at an iteration $k\ge 2b$). Note that the Lipschitz condition on $\gr f_i$ implies a Lipschitz condition and a bound on $\gr F$. In particular, denoting
$$L=\sum_{i=1}^mL_i,$$
we have for all $x,z\in\rn$,
$$\big\|\gr F(x)-\gr F(z)\big\|=\lf\|\sum_{i=1}^m\gr f_i(x)-\sum_{i=1}^m\gr f_i(z)\ri\|\le \sum_{i=1}^m\big\|\gr f_i(x)-\gr f_i(z)\big\| \le \sum_{i=1}^m L_i\|x-z\|=L\|x-z\|.\xdef\lipschitz{\lab}\eqnum\show{twoo}$$
As a special case, for $z=x^*$, where $x^*$ is the unique minimum of $F$, we have
$$\big\|\gr F(x_\ell)\big\|=\big\|\gr F(x_\ell)-\gr F(x^*)\big\|\le L\|x_\ell-x^*\|,\qquad \forall\ \ell\ge0.\xdef\gradbound{\lab}\eqnum\show{twoo}$$

We break down the proof of Prop.\ \propiapd\ in steps, first writing the iteration \incraggrproxt\ as a gradient iteration with errors, then carrying along the errors in a standard line of linear convergence analysis of gradient methods without errors, then bounding the errors, and finally using Lemma \lemmaone:

(a) We write the iteration \incraggrproxt\ as a gradient method with errors
$$x_{k+1}=x_k-\a\big(\gr F(x_k)+e_k\big),\xdef\relone{\lab}\eqnum\show{twoo}$$
where the error term $e_k$ is given by
$$e_k=\gr f_{i_k}(x_{k+1})-\gr f_{i_k}(x_{k})+\sum_{i\ne i_k}\big(\gr f_{i}(x_{\ell_i})-\gr f_{i}(x_{k})\big).\xdef\reltwo{\lab}\eqnum\show{twoo}$$
\smskip

{(b)} We relate the gradient error $e_k$ to the distance $\|x_k-x^*\|$ by verifying the relation
$$\|x_{k+1}-x^*\|^2=\|x_k-x^*\|^2-2\a\gr F(x_k)'(x_k-x^*)+\a^2\big\|\gr F(x_k)\big\|^2+E_k,\xdef\relfour{\lab}\eqnum\show{twoo}$$
where
$$E_k=\a^2\|e_k\|^2-2\a \big(x_k-\a\gr F(x_k)-x^*\big)'e_k.\xdef\relfive{\lab}\eqnum\show{twoo}$$
This is done by subtracting $x^*$ from both sides of Eq.\ \relone, norm-squaring both sides, and carrying out the straightforward calculation.

\smskip
{(c)} We use Eq.\ \relfive\ to bound $|E_k|$ according to
$$|E_k|\le \a^2\|e_k\|^2+2\a\|e_k\|\,\big\|x_k-x^*\big\|,\xdef\releight{\lab}\eqnum\show{twoo}$$
for all sufficiently small $\a$. In particular, from Eq.\ \relfive, we have 
$$|E_k|\le \a^2\|e_k\|^2+2\a\|e_k\|\,\big\|x_k-x^*-\a\gr F(x_k)\big\|,$$
and Eq.\ \releight\ is obtained from the preceding relation by using the inequality
$$\big\|x_k-x^*-\a\gr F(x_k)\big\|\le \|x_{k}-x^*\|.$$
which holds for $\a$ sufficiently small; this is a consequence of the fact that under the gradient Lipschitz assumption, a gradient iteration (with no error) reduces the distance to $x^*$ for $\a\in(0,1/L]$ (see e.g., [Ber15], Prop.\ 6.1.6).

\smskip
{(d)} We use the strong convexity assumption
$$\big(\nabla
F(x)-\nabla F(y)\big)\tr (x-y) \geq \s\|x-y\|^2, \qquad \forall\
x,y\in\rn,\xdef\relsix{\lab}\eqnum\show{twoo}
$$
where $\s$ is the coefficient of strong convexity and the Lipschitz condition \lipschitz, to invoke the relation
$$\gr F(x_k)'(x_k-x^*)\ge {\s L\over \s+L}\|x_k-x^*\|^2+{1\over \s+L}\big\|\gr F(x_k)\big\|^2;\xdef\relsix{\lab}\eqnum\show{twoo}$$
see e.g., [Nes14], Th.\ 2.1.22, or [Ber15], Prop.\ 6.1.9(b). This will be used to bound the term $\gr F(x_k)'(x_k-x^*)$ of Eq.\ \relfour.

\smskip
{(e)} We show that for $\a\le {2\over \s+L}$, we have
$$\|x_{k+1}-x^*\|^2\le \lf(1-2\a{\s L\over \s+L}\ri)\|x_k-x^*\|^2+|E_k|.\xdef\relseven{\lab}\eqnum\show{twoo}$$
 In particular, using the relations \relfour\ and \relsix, we have 
 $$\eqalign{\|x_{k+1}-x^*\|^2&\le \|x_k-x^*\|^2-2\a\lf({\s L\over \s+L}\|x_k-x^*\|^2+{1\over \s+L}\big\|\gr F(x_k)\big\|^2\ri)+\a^2\big\|\gr F(x_k)\big\|^2+|E_k|\cr
 &\le \lf(1-2\a{\s L\over \s+L}\ri)\|x_k-x^*\|^2+\a\lf(\a-{2\over \s+L}\ri)\big\|\gr F(x_k)\big\|^2+|E_k|,\cr}$$
from which Eq.\ \relseven\ follows.

\smskip

{(f)} We prove that the error $e_k$ is proportional to the stepsize $\a$, and to the maximum distance  of the iterates from $x^*$ over the past $2b$ iterates:
$$\|e_k\|\le O(\a)\max_{k-2b\le \ell\le k}\|x_\ell-x^*\|.\xdef\relthree{\lab}\eqnum\show{twoo}$$
This is straightforward, using the Lipschitz assumption on $\gr f_i$ and the bound \gradbound\ on $\gr F$. 

In particular, from Eq.\ \reltwo, we have
$$\eqalign{\|e_k\|&\le \big\|\gr f_{i_k}(x_{k+1})-\gr f_{i_k}(x_k)\big\|+\sum_{i\ne i_k} \big\|\gr f_i(x_{\ell_i})- \gr f_i(x_{k})\big\|\cr
&\le L_{i_k}\|x_{k+1}-x_k\|+\sum_{i\ne i_k} L_i\| x_{k}-x_{\ell_i}\|\cr
&\le L_{i_k}\|x_{k+1}-x_k\|+\sum_{i\ne i_k} L_i\big(\| x_{k}-x_{k-1}\|+\cdots+\| x_{\ell_i+1}-x_{\ell_i}\|\big).\cr}\xdef\relthreet{\lab}\eqnum\show{twoo}$$
Moreover from Eqs.\ \gradbound\ and \relone,
$$\|x_{\ell+1}-x_\ell\|=\a\big\|\gr F(x_\ell)\big\|+\a\|e_\ell\|\le \a L\|x_\ell-x^*\|+\a\|e_\ell\|,\qquad \forall\ \ell\ge0.\xdef\relthreett{\lab}\eqnum\show{twoo}$$
Using this relation for $\ell$ in the range $[k-b,k]$ in Eq.\ \relthreet, we obtain
$$(1-\a L_{i_k})\|e_k\|\le O(\a)\lf(\sum_{\ell=k-b}^k\|x_\ell-x^*\|+\sum_{\ell=k-b}^{k-1}\|e_\ell\|\ri),$$
where for $p\ge 1$, we generically use $O(\a^p)$ to denote any function of $\a$ such that for some scalar $\g>0$, we have $\big|O(\a^p)\big|\le \g\a^p$ for all $\a$ in some bounded open interval containing the origin. Thus,
$$\|e_k\|\le O(\a)\lf(\sum_{\ell=k-b}^k\|x_\ell-x^*\|+\sum_{\ell=k-b}^{k-1}\|e_\ell\|\ri).\xdef\ekbound{\lab}\eqnum\show{twoo}$$
From Eq.\ \reltwo, we also have
$$\eqalign{\|e_\ell\|&\le L_{i_\ell}\|x_{\ell+1}-x_\ell\|+\sum_{i\ne i_\ell}L_i\|x_\ell-x_{\ell_i}\|\cr
&\le L\lf(\|x_{\ell+1}-x^*\|+\|x_\ell-x^*\|+\sum_{i\ne i_\ell}L_i\big(\|x_\ell-x^*\|+\|x_{\ell_i}-x^*\|\big)\ri).\cr}\xdef\elbound{\lab}\eqnum\show{twoo}$$
Since for $\ell$ in the range $[k-b,k-1]$, $\ell_i$ lies in the range $[k-2b,k-1]$, it follows that
$$\|e_\ell\|\le c\max_{k-2b\le \ell\le k}\|x_\ell-x^*\|,\qquad \forall\ \ell\in [k-b,k-1],$$
where $c$ is some constant that is independent of $k$ and $\ell$. Combining this with Eq.\ \ekbound, we obtain Eq.\ \relthree. 

\smskip

{(g)} We use Eqs.\ \releight, \relseven, and  \relthree\ to obtain
$$\|x_{k+1}-x^*\|^2\le \lf(1-2\a{\s L\over \s+L}\ri)\|x_k-x^*\|^2+O(\a^2)\max_{k-2b\le \ell\le k}\|x_\ell-x^*\|^2.\xdef\relnine{\lab}\eqnum\show{twoo}$$
 In particular, the two terms bounding $|E_k|$ in Eq.\ \releight\ are $\a^2\|e_k\|^2$ and $\a\|e_k\|\,\big\|x_k-x^*\big\|$, which in view of Eq.\ \relthree\ are bounded by terms that are $O(\a^4)$ and $O(\a^2)$ times $\max_{k-2b\le \ell\le k}\|x_\ell-x^*\|^2$, respectively.

\smskip
{(h)} We use Eq.\ \relnine\ and Lemma \lemmaone, with $d=2b$, $\b_k=\|x_k-x^*\|^2$, $p=1-2\a{\s L\over \s+L}$, and $q=O(\a^2)$, so that $p+q<1$ for sufficiently small $\a$.  This shows that $\sqrt{\b_k}=\|x_k-x^*\|$ converges  linearly to 0, and completes the proof.
\qed

\old{
In the case where there is an error $w_k$ in the calculation of the delayed gradients as per Eq.\ \erroriter, and $w_k$ satisfies  for some $\b>0$,
$$\|w_k\|\le \b\sum_{\ell =\max\{0,k-b\}}^{k+1}\|x_\ell-x_{\ell-1}\|,\qquad\forall\ k\ge 0,\xdef\delaycond{\lab}\eqnum\show{twoo}$$
the result can be shown with a simple modification of the preceding proof.
By redefining the error  $e_k$ of Eq.\ \reltwo\ as
$$e_k=\gr f_{i_k}(x_{k+1})-\gr f_{i_k}(x_{k})+\sum_{i\ne i_k}\big(\gr f_{i}(x_{\ell_i})-\gr f_{i}(x_{k})\big)+w_k,$$
steps (b)-(e) of the proof  go through. Moreover, Eq.\ \relthreet\ now becomes
$$\|e_k\|\le L_{i_k}\|x_{k+1}-x_k\|+\sum_{i\ne i_k} L_i\big(\| x_{k}-x_{k-1}\|+\cdots+\| x_{\ell_i+1}-x_{\ell_i}\|\big)+\|w_k\|,\xdef\relthreetto{\lab}\eqnum\show{twoo}$$
while Eq.\ \relthreett\ still holds, i.e.,
$$\|x_{\ell+1}-x_\ell\|=\a\big\|\gr F(x_\ell)\big\|+\a\|e_\ell\|\le \a L\|x_\ell-x^*\|+\a\|e_\ell\|,\qquad \forall\ \ell\ge0.\xdef\relthreettt{\lab}\eqnum\show{twoo}$$
Combining Eqs.\ \delaycond\ and \relthreetto\ with Eq.\ \relthreettt\ for $\ell=k$, we obtain
$$\eqalign{(1-\a \big(L_{i_k}+\b)\big)\|e_k\|\le \a L\|x_k-x^*\|&+\sum_{i\ne i_k} L_i\big(\| x_{k}-x_{k-1}\|+\cdots+\| x_{\ell_i+1}-x_{\ell_i}\|\big)\cr
&+\b\sum_{\ell =\max\{0,k-b\}}^{k}\|x_\ell-x_{\ell-1}\|.\cr}$$
Applying  Eq.\ \relthreettt\ for $\ell=k-1,\ldots,\ell_i$, the assumption \delaycond, we still obtain Eq.\ \ekbound, i.e.,
$$\|e_k\|\le O(\a)\lf(\sum_{\ell=k-b}^k\|x_\ell-x^*\|+\sum_{\ell=k-b}^{k-1}\|e_\ell\|\ri).\xdef\ekbound{\lab}\eqnum\show{twoo}$$
while
in place of Eq.\ \elbound, we have
$$\|e_\ell\|\le L\lf(\|x_{\ell+1}-x^*\|+\|x_\ell-x^*\|+\sum_{i\ne i_\ell}L_i\big(\|x_\ell-x^*\|+\|x_{\ell_i}-x^*\|\big)\ri)+\|w_\ell\|.$$
Combining the last two relations, using the assumption \delaycond,  and collecting terms, we see that in place of Eq.\ \relthree, we obtain
$$\|e_k\|\le O(\a)\max_{k-2b\le \ell\le k}\|x_\ell-x^*\|+O(\a)\max_{k-b\le \ell\le k-1}\|w_\ell\|\le O(\a)\max_{k-2b\le \ell\le k}\|x_\ell-x^*\|,$$
 where the last inequality follows using the assumption \delaycond. Thus we have recovered Eq.\ \relthree, and the remainder of the proof goes through as before.
 }

\subsubsection{Convergence Rate Comparison for Small Stepsizes}

\pn Note that Eq.\ \relnine\ provides a more refined rate of convergence estimate. While this estimate is not very precise,
because of the second order term on the right in Eq.\ \relnine, it shows that the ratio 
$${\s L\over \s+L}={L\over 1+L/\s}$$
where $L=\sum_{i=1}^mL_i$ and $\s$ is the coefficient of strong convexity, plays an important role, and in particular the convergence rate is improved when the ``condition number" $L/\s$ is small. The role of the ratio $L/\s$ in determining the convergence rate of gradient methods (without error) is well-known; see e.g., [Nes04], [Ber15]. 

Convergence rate estimates like the one of Eq.\ \relnine\ can also be similarly derived for IAG (as shown in [GOP15]), and for the standard nonincremental gradient method [for which the error term $|E_k|$ in Eq.\ \relseven\ is equal to 0]. These estimates, to first order [i.e., after  neglecting the second order term in the right-hand side of Eq.\ \relnine], are identical for IAP, IAG, and for the standard nonincremental gradient method. This suggests that for very small values of $\a$, IAP and IAG perform comparably, while the  nonincremental gradient method performs much worse because it requires $m$ times as much overhead per iteration to calculate the full gradient of the cost function. 

\vskip-1pc

\section{Nonquadratic Incremental Proximal and Augmented Lagrangian\hfill\break Methods}

\pn The augmented Lagrangian methods of Section 2 apply to linear equality constrained problems for which the multiplier vector $\l$ is unconstrained. This allows the application of the linear convergence result of Prop.\ \propiapd. We will now consider convex inequality constraints, whose multipliers must be nonnegative. As a result the dual problem involves an orthant constraint, and the linear convergence result of Prop.\ \propiapd\ does not apply. Unfortunately, when there is an orthant constraint [i.e., $X=\{x\mid x\ge0\}$ instead of $X=\rn$ in Eq.\ \incraggrprox], the proof of Prop.\ \propiapd\ breaks down because the critical inequality \gradbound\ fails. In fact, to our knowledge, a linear convergence rate result for the IAG method \iagiter\ applied with an orthant constraint is not currently available. Moreover, the convergence of the augmented Lagrangian-like methods discussed in Section 2 has been analyzed only for the equality-constrained case. In this section we will try to address this difficulty by using a different (nonquadratic) proximal approach.

In particular, we will introduce incremental augmented Lagrangian methods for convex inequality constraints, where the quadratic penalty in the augmented Lagrangian is replaced by a suitable nonquadratic penalty. One of our objectives is to develop  linearly convergent methods that can exploit separability, similar to the ones of Section 2. A second objective is to develop corresponding dual linearly convergent incremental aggregated gradient and proximal methods for differentiable minimization subject to nonnegativity constraints.

\subsubsection{Nonquadratic Augmented Lagrangian Methods for Inequality Constraints}

\pn Consider the convex programming problem
$$\eqalign{&\hbox{minimize\ \ \ }H(y)\cr
&\hbox{subject to\ \ }y\in Y,\ \ \ \ \ G_j(y)\le 0,\ \ j=1,\ldots,r,\cr}\xdef\genconstrproblem{\lab}\eqnum\show{oneo}
$$
where $H:\rn\mapsto(-\infty,\infty)$ and $G_j:\rn\mapsto(-\infty,\infty)$ are convex functiona, and $Y$ is a convex set. The corresponding dual problem is
$$\eqalign{&\hbox{maximize\ \ \ }Q(\m)\cr
&\hbox{subject to\ \ }\m\ge 0,\cr}\xdef\genconstrproblem{\lab}\eqnum\show{oneo}
$$
where $Q:\re^r\mapsto[-\infty,\infty)$ is the concave function of the multiplier vector $\m=(\m^1,\ldots,\m^r)$, given by
$$Q(\m)=\inf_{y\in Y}\lf\{H(y)+\sum_{j=1}^r\m^jG_j(y)\ri\},\qquad \m\in
\re^r.\xdef\dualfunctq{\lab}\eqnum\show{oneo}$$

We will apply an augmented Lagrangian method, first proposed by Kort and Bertsekas [KoB72], and further developed in a number of subsequent works, including the monograph [Ber82] (Chapter 5).  The method makes use of a  nonquadratic penalty function  $\psi:\re\mapsto\re$ with the following properties:
\nitem{(i)} $\psi$ is twice differentiable and $\gr^2\psi(t)>0$ for all $t\in\re$, 
\nitem{(ii)} $\psi(0)=0$, $\gr\psi(0)=1$,
\nitem{(iii)} $\lim_{t\to-\infty}\psi(t)>-\infty$, 
\nitem{(iv)} $\lim_{t\to-\infty}\gr\psi(t)=0$ and
$\lim_{t\to\infty}\gr\psi(t)=\infty$.
\smskip
\pn The most common and interesting special case is the exponential
$$\psi(s)	= \exp(s) - 1,\qquad s\in\re.\xdef\exppen{\lab}\eqnum\show{oneo}$$
The corresponding {\it exponential augmented Lagrangian method\/} and its dual, a proximal algorithm known as the {\it entropy minimization algorithm\/}, has been analyzed first in [KoB72] and [Ber82], and then by Tseng and Bertsekas [TsB93]. Related classes of methods, which also contain the exponential and entropy methods as special cases, were proposed and analyzed later by Iusem, Svaiter, and Teboulle [IST94]; see also the survey by Iusem [Ius99], which contains followup work and many references.

The augmented Lagrangian algorithm corresponding to $\psi$ and  problem \genconstrproblem\ maintains multipliers $\m_k^j>0$, $j=1,\ldots,r$, for the inequality constraints, and consists of finding
$$y_{k+1}\in\arg\min_{y\in Y}\lf\{H(y)+\sum_{j=1}^r{\m_k^j\over \a_k^j}\psi\big(\a_k^j G_j(y)\big)\ri\},\xdef\minaugmnq{\lab}\eqnum\show{oneo}
$$
where $\a_k^j>0$, $j=1,\ldots,r$, are penalty parameters, followed by the multiplier iteration
$$\m_{k+1}^j=\m_k^j\gr \psi\big(a_k^jG_j(y_{k+1})\big),\qquad j=1,\ldots,r.\xdef\multiternq{\lab}\eqnum\show{oneo}$$
Alternatively and equivalently, based on the Fenchel duality theorem, one may show that the multiplier iteration can be written in the  proximal form
$$\m_{k+1}\in \arg\max_{\m\in\re^r}\lf\{Q(\m)-\sum_{j=1}^r{\m_k^j\over \a_k^j}\psi^\star\lf({\m^j\over \m_k^j}\ri)\ri\},\xdef\multprox{\lab}\eqnum\show{oneo}$$
where $Q$ is the dual function given by Eq.\ \dualfunctq,
and $\psi^\star$ is the convex conjugate of $\psi$. 

To see the equivalence of the expressions \multiternq\ and \multprox, let us write
$$u_{k+1}^j=G_j(y_{k+1}),\qquad j=1,\ldots,r,$$
 and note that the augmented Lagrangian minimization \minaugmnq\ yields
$$u_{k+1}\in \arg\min_{u=(u^1,\ldots,u^r)\in\re^r}\lf\{P(u)+\sum_{j=1}^r{\m_k^j\over \a_k^j}\psi(\a_k^j u^j)\ri\},\xdef\primalfenchel{\lab}\eqnum\show{oneo}$$
where $P$ is the primal function
$$P(u)=\inf_{y\in Y,\ G_j(y)\le u^j,\, j=1,\ldots,r}H(y).$$
Then the minimization in Eq.\ \primalfenchel\ is the Fenchel dual to the maximization 
 \multprox. By applying the Fenchel duality theorem, we have that the maximizing vector in Eq.\ \multprox\ is equal to the gradient
 $$\gr \lf(\sum_{j=1}^r{\m_k^j\over \a_k^j}\psi(\a_k^j u^j)\ri)\Bigg|_{u=u_{k+1}},$$
 so it is given by the formula \multiternq.

Note that while the dual problem is to maximize $Q(\m)$ subject to $\m\ge0$, the proximal maximization \multprox\ is unconstrained. The reason is that the conjugate $\psi^\star$ takes the value $\infty$ outside the nonnegative orthant, and has the character of a barrier function within the nonnegative orthant. As an example, for the exponential function \exppen\ the conjugate is the entropy function
$$\psi^\star(t)=\cases{t\big(\ln(t)-1\big)+1&if $t>0$,\cr
1&if $t=0$,\cr
\infty&if $t<0$.\cr}\xdef\entropyprox{\lab}\eqnum\show{oneo}$$

An important advantage of the nonquadratic augmented Lagrangian method versus its quadratic counterpart, is that it leads to twice differentiable augmented Lagrangians. This advantage also carries over to the incremental augmented Lagrangian methods to be presented next.

\subsubsection{Nonquadratic Incremental Augmented Lagrangian Methods for Inequality Constraints}

\pn Consider now the separable constrained optimization problem
$$\eqalign{&\hbox{minimize\ \ \ }{\sum_{i=1}^mh_i(y^i)}\cr
&\hbox{subject to\ \ }y^i\in Y_i, \ \ i=1,\ldots,m,\ \ \ \ \ \sum_{i=1}^mg_{ji}(y^i)\le 0,\cr}\xdef\thrsc{\lab}\eqnum\show{oneo}
$$
where $h_i$ and $g_{ji}$ are convex real-valued functions, and $Y_i$ are convex sets. Similar to the development of Section 2, the corresponding incremental aggregated augmented Lagrangian method, which parallels IAAL, maintains a vector $\m_k>0$ and operates as follows.

\texshopbox
{{\bf \pn Incremental Aggregated Augmented Lagrangian Iteration for Inequalities  (IAALI)}
\smskip
\pn Select a component index $i_k$, and update the single component $y^{i_k}$ according to
$$y_{k+1}^{i_k}\in\arg\min_{y^{i_k}\in Y_{i_k}}\lf\{h_{i_k}(y^{i_k})+\sum_{j=1}^r{\m_k^j\over \a_k^j}\psi\lf(\a_k^j \lf( g_{ji_k}(y^{i_k})+\sum_{i\ne i_k} g_{ji}(y_{\ell_i}^{i})\ri) \ri)\ri\},\xdef\agrauglagrmintnonq{\lab}\eqnum\show{twoo}$$
while keeping the others unchanged, $y^i_{k+1}=y^i_k$ for all $i\ne i_k$. Then update $\m$ according to 
$$\m^j_{k+1}=\m_k^j\gr \psi\lf(a_k^j\lf(g_{ji_k}(y^{i_k}_{k+1})+\sum_{i\ne i_k}^mg_{ji}(y^i_{\ell_i})\ri)\ri),\quad j=1,\ldots,r.\xdef\agrmultitertnonq{\lab}\eqnum\show{twoo}$$
}

Note that the minimization \agrauglagrmintnonq\ is of low dimension, but involves the nonquadratic penalty function $\psi$. Thus even when the component $y^{i_k}$ is one-dimensional, this minimization will likely require some form of iterative line search. Note also that the update formula \agrmultitertnonq\ can equivalently be written as
$$\m_{k+1}\in \arg\max_{\m\in\re^r}\lf\{q_{i_k}(\m)+\sum_{i\ne i_k}\gr q_{i}(\m_{\ell_i})'(\m-\m_k)-\sum_{j=1}^r{\m_k^j\over \a_k^j}\psi^\star\lf({\m^j\over \m_k^j}\ri)\ri\},\xdef\multproxnq{\lab}\eqnum\show{oneo}$$
where $q_i$ are the dual function components, given by 
$$q_i(\m)=\inf_{y^i\in Y^i}\lf\{h_i(y^i)+\sum_{j=1}^r\m^jg_{ij}(y^i)\ri\},\qquad \m\in
\re^r,\qquad i=1,\ldots,m.$$
The form \multproxnq\ of the method can be viewed as an incremental aggregated proximal method for maximizing $Q(\m)=\sum_{i=1}^mq_i(\m)$ over $\m\ge0$, where
$$q_i(\m)=\inf_{y^i\in Y_i}\lf\{h_i(y^i)+\sum_{j=1}^r\m^j g_{ji}(y^i)\ri\},\qquad i=1,\ldots,m;\eqnum$$
cf.\ Eq.\ \lagrangianmin. The convergence properties of the IAALI and the corresponding incremental aggregated proximal method \multproxnq\ for solving the dual problem
$$\eqalign{&\hbox{maximize\ \ \ }{\sum_{i=1}^m q_i(\m)}\cr
&\hbox{subject to\ \ }\m\ge0,\cr}
$$
are interesting research subjects, as we will now discuss.

\subsubsection{Nonquadratic Incremental Aggregated Proximal Algorithm for Nonnegativity Constraints}

\pn Consider the minimization problem
$$\eqalign{\hbox{\rm minimize}\quad &F(x)\; {\buildrel\rm def\over=} \; \sum_{i=1}^mf_i(x)\cr
\hbox{\rm subject to\ \ }
&x\ge0,\cr}\xdef\nonnegprob{\lab}\eqnum\show{twoo}$$
where $f_i:\rn\mapsto\re$, $i=1,\ldots,m$,  are convex real-valued functions. When translated to this minimization context, the algorithm \multproxnq\ maintains a vector $x_k>0$ that is updated as follows.

\texshopbox{{\bf \pn Nonquadratic Incremental Aggregated Proximal Iteration for $X=\{x\mid x\ge0\}$}
\smskip
\pn Select a component index $i_k$, and obtain $x_{k+1}$ as
$$x_{k+1}\in\arg\min_{x\in \rn}\lf\{f_{i_k}(x)+\sum_{i\ne i_k}\gr f_{i}(x_{\ell_i})'(x-x_k)+\sum_{j=1}^n{x_k^j\over \a_k^j}\psi^\star\lf({x^j\over x_k^j}\ri)\ri\}.\xdef\additiveproxgen{\lab}\eqnum\show{twoo}$$
}

The analysis of the convergence properties of this algorithm is beyond the scope of this paper, and will be the subject of a separate publication.  In particular, it is interesting to investigate the linear convergence of the method \additiveproxgen\ when the parameters $a_k^j$ are constant (but sufficiently small), under the appropriate  Lipschitz continuity and strong convexity assumptions, similar to Prop.\ \propiapd. Note that by differentiating the cost function in the  minimization of Eq.\ \additiveproxgen, we obtain the optimality condition, which can be written as
$$\gr f_{i_k}(x_{k+1})+\sum_{i\ne i_k} \gr f_i(x_{\ell_i})+\pmatrix{{1\over \a_k^1}\gr \psi^\star\lf({x^1_{k+1}\over x_k^1}\ri)\cr
\vdots\cr
{1\over \a_k^m}\gr \psi^\star\lf({x^m_{k+1}\over x_k^m}\ri)\cr}=0.\xdef\genproxiter{\lab}\eqnum\show{twoo}$$
This expression may be used in the line of proof of Section 3 in place of the corresponding formula \incraggrproxfif\
for the unconstrained IAP algorithm \incraggrproxfif, which can be written in the form
$$\gr f_{i_k}(x_{k+1})+\sum_{i\ne i_k}\gr f_i(x_{\ell_i})+{x_{k+1}-x_k\over \a}=0.\xdef\quadproxiter{\lab}\eqnum\show{twoo}$$
When $\psi$ (and hence also $\psi^\star$) is quadratic and $\a_k^j\equiv\a$, the two preceding formulas coincide. However, contrary to iteration \quadproxiter, the iteration \genproxiter\ preserves the strict positivity of the iterates ($x_k>0$ for all $k$), and addresses the orthant-constrained problem \nonnegprob.

\subsubsection{Entropy-Based Incremental Aggregated Proximal Algorithm for Nonnegativity Constraints}

\pn For an illustration  of the algorithm \additiveproxgen, consider the special case where $\psi$ is the exponential function  and $\psi^\star$ is the entropy function, so that  
$$\psi(s)	= \exp(s) - 1,\quad\  \psi^\star(t)=\cases{t\big(\ln(t)-1\big)+1&if $t>0$,\cr
1&if $t=0$,\cr
\infty&if $t<0$,\cr}\quad\  \gr \psi^\star(t)=\cases{\ln(t)&if $t>0$,\qquad \cr
\hbox{does not exist}&if $t\le 0$,\cr}$$
[cf.\ Eqs.\ \exppen\ and \entropyprox].  Then by using a constant stepsize $\a^j$ for each coordinate, Eq.\ \genproxiter\ takes the form
$$\ln\lf({x^j_{k+1}\over x_k^j}\ri) =-\a^j \lf({\partial f_{i_k}(x_{k+1})\over \partial x^j}+\sum_{i\ne i_k} {\partial f_{i}(x_{\ell_i})\over \partial x^j}\ri),\qquad j=1,\ldots,n,\xdef\entropyiap{\lab}\eqnum\show{twoo}$$
where $i_k$ is the component index selected for iteration $k$. We can write this iteration as
$$\ln\lf({x^j_{k+1}\over x_k^j}\ri) =-\a^j \lf({\partial F(x_k)\over \partial x^j}+e_k^j\ri),\xdef\logiter{\lab}\eqnum\show{twoo}$$
where $e_k=(e_k^1,\ldots,e_k^n)$ is the error vector
$$e_k=\gr f_{i_k}(x_{k+1})-\gr f_{i_k}(x_{k})+\sum_{i\ne i_k}\big(\gr f_{i}(x_{\ell_i})-\gr f_{i}(x_{k})\big),\xdef\graderror{\lab}\eqnum\show{twoo}$$
that played an important role in the proof of  Prop.\ \propiapd\ [cf.\ Eq.\ \reltwo].

We will use the line of analysis of Section 3 to speculate about the linear convergence of iteration \entropyiap\ and its equivalent form \logiter-\graderror. Assume that the minimum $x^*$ satisfies the strict complementary slackness condition
$${\partial F(x^*)\over \partial x^j}>0,\qquad \forall\ j\in J^0,\xdef\strictcompl{\lab}\eqnum\show{twoo}$$
where $J^0=\big\{j\mid (x^j)^*=0\big\}$, and speculate on the behavior of $\{x_k\}$ in a small neighborhood around $x^*$.

Consider first the iterates $x_k^j$, $ j\in J^0$, in a small neighborhood around $x^*$. We note that the errors $e_k^j$ of Eq.\ \graderror\ are near 0 and by Eq.\ \strictcompl, are negligible relative to the gradient components ${\partial F(x_k)\over \partial x^j}$, for all $j\in J^0$. In view of the form of iteration \logiter\ and the condition \strictcompl, the logarithms $\ln(x_{k+1}^j/x_k^j)$, $ j\in J^0$, are negative, and hence the ratios $x_{k+1}^j/x_k^j$, $ j\in J^0$, are within $[0,1)$, so the sequences $\{x_k^j\}$, $ j\in J^0$, are linearly decreasing towards 0. 

Consider next the iterates $x_k^j$, $ j\notin J^0$, in a small neighborhood around $x^*$. They are close to the corresponding positive numbers $(x^j)^*$, $ j\notin J^0$,  and they are iterated according to
$$\ln({x^j_{k+1}) =\ln(x_k^j}) -\a^j \lf({\partial F(x_k)\over \partial x^j}+e_k^j\ri),\qquad j\notin J^0,\xdef\lniter{\lab}\eqnum\show{twoo}$$
[cf.\ Eq.\ \logiter]. This looks like an incremental aggregated gradient iteration in the logarithms $\ln(x^j)$, $j\notin J^0$. Indeed by making the transformation of variables $z^j=\ln(x^j)$, $j=1,\ldots,n$, for $x^j>0$, and introducing the function
$$H(z^1,\ldots,z^n)=F\big(\exp({z^1}),\ldots,\exp({z^n})\big),$$
and its gradient, which is related to the gradient of $F$ through the relation
$${\partial H(z)\over \partial z^j}=\exp({z^j}){\partial F\big(\exp({z^1}),\ldots,\exp({z^n})\big)\over \partial x^j}=x^j{\partial F(x)\over \partial x^j},\qquad j=1,\dots,n,$$
we see that the iteration \lniter\ can be written as
$$z_{k+1}^j=z_k^j-{\a^j \over x_k^j}\lf({\partial H(z_k)\over \partial z^j}\ri)+\a^je_k^j,\qquad j\notin J^0,$$
where $x_k^j=\exp({z_k^j})$.
Thus, neglecting the effect of the coordinates $x^j$, $j\in J^0$, that are fast diminishing to 0, the iteration behaves like the IAP method restricted to the space of the coordinate logarithms $z^j=\ln(x^j)$, $j\notin J^0$, with coordinate-dependent stepsizes ${\a^j \over x_k^j}$ that are close to the positive constants ${\a^j \over (x^j)^*}$, $j\in J^0$, for $x_k$ near $x^*$. 

By combining the preceding argument with the proof of  Prop.\ \propiapd, we can show that the method converges to $x^*$ locally, i.e., when started sufficiently close to $x^*$, assuming  the strict complementarity condition \strictcompl, and the appropriate stepsize, Lipschitz continuity, and strong convexity conditions. The proof is long and will be deferred to a future publication. Moreover, for $j\notin J^0$, $\big\{\ln(x_k^j)\big\}$ converges to $\ln\big((x^j)^*\big)$ linearly, while for $j\in J^0$, $\{x_k^j\}$ also converges to $(x^j)^*$ linearly. However, a more sophisticated argument is needed to show global and linear convergence of $\{x_k\}$ to $x^*$, by combining the line of proof of  Prop.\ \propiapd\ with the existing convergence proofs of the entropy minimization algorithm and its dual, the exponential method of multipliers.

\subsubsection{Entropy-Based Incremental Aggregated Gradient Algorithm for Nonnegativity Constraints}

\pn Finally let us note the analog of the IAG method for nonnegativity constraints. In analogy with Eq.\ \entropyiap\ it has the form
$$\ln\lf({x^j_{k+1}\over x_k^j}\ri) =-\a^j \sum_{i=1}^m {\partial f_{i}(x_{\ell_i})\over \partial x^j},\qquad j=1,\ldots,n,$$
or equivalently
$$x^j_{k+1}=x_k^j \exp\lf(-\a^j \sum_{i=1}^m {\partial f_{i}(x_{\ell_i})\over \partial x^j}\ri),\qquad j=1,\ldots,n,\xdef\entropyiag{\lab}\eqnum\show{twoo}$$
[the difference from Eq.\ \entropyiap\ is the use of ${\partial f_{i_k}(x_{\ell_{i_k}})\over \partial x^j}$ in place of ${\partial f_{i_k}(x_{k+1})\over \partial x^j}$]. This iteration should be compared with the IAS method \distrincraggrproxproj, for the case where the functions $f_i$ are differentiable, and the stepise $\a_k$ is a constant $\a$:
$$x_{k+1}=\lf[x_k-\a\sum_{i=1}^m \gr f_i(x_{\ell_i})\ri]^+,\xdef\quadiag{\lab}\eqnum\show{twoo}$$
where $[\cdot]^+$ denotes projection onto the nonnegative orthant. We may view the method \quadiag\ as the constrained version of the IAG method \iagiter\ with constant stepsize for which, however, no linear convergence proof is presently available.\footnote{\dag}{\ninepoint  A local linear convergence result  for the constrained IAG method \quadiag\ is possible, assuming  the strict complementarity condition \strictcompl. In particular, it can be shown that there is a sphere centered at  $x^*$ such that if $x_0$ belongs to that sphere, then the sequence generated by iteration \quadiag\ stays within that sphere and converges linearly to $x^*$. The idea of the proof is that after the first iteration, all the iterates satisfy $x_k^j=0$ for all indices $j\in J^0$, so the method essentially reduces to the IAG method in the space of variables $x^j$, $j\notin J^0$.}

\old{This iteration can also be written as
$$x_{k+1}\in\arg\min_{x\in \rn}\lf\{\gr f_{i_k}(x_{k})'(x-x_k)+\sum_{i\ne i_k}\gr f_{i}(x_{\ell_i})'(x-x_k)+\sum_{j=1}^n{x_k^j\over \a_k^j}\psi^\star\lf({x^j\over x_k^j}\ri)\ri\},\old{\eqnum\show{twoo}}$$
which is a more general form that is valid for a general function $\psi$.
When $\psi$ (and hence also $\psi^\star$) is quadratic and $\a_k^j\equiv\a$, it becomes similar to the IAG iteration \iagiter. However, like the iteration \genproxiter, it preserves the strict positivity of the iterates ($x_k>0$ for all $k$), and addresses the orthant-constrained problem \nonnegprob.}

The iteration \entropyiag\ may also be viewed as an incremental version of the mirror descent method; see Beck and Teboulle [BeT03], the surveys by Juditsky and Nemirovski [JuN11a], [JunN11b], and the references quoted there, and the author's presentation in [Ber15], Section 6.6. Using similar arguments to the case of iteration \entropyiap, we can show that the iteration \entropyiag\ converges linearly to $x^*$, when started sufficiently close to $x^*$, assuming  the strict complementarity condition \strictcompl, and the appropriate constant stepsize, and other conditions.
Note that the iteration  \entropyiag\ may be implemented more conveniently than the proximal iteration \additiveproxgen, as it does not require a proximal minimization. However,  the iteration  \entropyiag\  is not suitable as the basis for the development of an incremental augmented Lagrangian method, such as IAALI [cf.\ Eqs.\ \agrauglagrmintnonq-\agrmultitertnonq].

A final comment relates to the choice of the stepsizes $\a^j$ in iteration \entropyiag. For the coordinates that are bounded away from 0 (i.e., for $j\notin J^0$) we have asymptotically $\sum_{i=1}^m {\partial f_{i}(x_{\ell_i})\over \partial x^j}\approx 0$, so from a Taylor expansion of the exponential in Eq.\ \entropyiag, we obtain
$$x^j_{k+1}=x_k^j \lf(1+\lf(-\a^j \sum_{i=1}^m {\partial f_{i}(x_{\ell_i})\over \partial x^j}\ri)+{1\over 2}\lf(-\a^j \sum_{i=1}^m {\partial f_{i}(x_{\ell_i})\over \partial x^j}\ri)^2+\cdots \ri).$$
By discarding the second and higher order terms for $j\notin J^0$, we see that approximately,
$$x^j_{k+1}\approx x_k^j-\a^jx_k^j\sum_{i=1}^m {\partial f_{i}(x_{\ell_i})\over \partial x^j},\qquad j\notin J^0.$$
This suggests scaling the stepsizes $\a^j$ for $j\notin J^0$, so that $\a^j$ is inversely proportional to the optimal value $(x^j)^*$. On the other hand, for $j\in J^0$, it makes sense to  choose $\a^j$ large (subject to a positive lower bound) in order to accelerate the convergence of $x_k^j$ to $(x^j)^*=0$. Thus a reasonable heuristic is to set
$$\a^j={\a\over \max\{\bar x^j,\,\d\}},\qquad j=1,\ldots,n,$$
where $\bar x^j$ is an estimate  for the optimal coordinate value $(x^j)^*$, $\a$ is some positive scalar, which corresponds to the stepsize of the constrained IAG iteration \quadiag, and $\d$ is a small positive constant. One may also consider updating the values $\a^j$ in the course of the algorithm, as better estimates $\bar x^j$ are obtained.

\vskip-1pc

\section{Concluding Remarks}
\vskip-0.5pc

\pn In this paper we have proposed IAP, an incremental aggregated proximal method, and we have shown that under favorable assumptions, it attains a linear convergence rate, using a constant (but sufficiently small) stepsize. The application of this method in a dual context, to separable constrained optimization problems, yields the IAAL method, an incremental augmented Lagrangian method that preserves and exploits the separable structure. The principal difference of our method relative to the several alternative augmented Lagrangian-based proposals, is its incremental character and its high update frequency of the multiplier $\l_k$; the alternative methods, except Algorithm 1 of [WHM13] and the one of [RoT15], but including the proper version of ADMM for separable problems,  update all the primal variables $y^i$, $i=1,\ldots,m$, simultaneously rather than sequentially, so they are not incremental in nature. Moreover, the alternative methods update the multipliers $m$ times less frequently than IAAL. A systematic computational comparison of our methods with the nonincremental alternatives will be helpful in clarifying what advantages our incremental approach may hold.  
 
There are several analytical issues relating to the IAAL method, which require further investigation. For example a more refined convergence rate analysis may point the way to adaptive stepsize adjustment schemes, and/or forms of scaling based on second derivatives of the cost function and the matrices $A_i$. There are analyses of this type for ADMM; see the paper by Giselsson  and Boyd [GiB15], and the references cited there. Another possibility is to use a momentum term in the updating formula for the multiplier $\l$. A third possibility is to control the degree of incrementalism by ``batching" multiple augmented Lagrangian iterations involving multiple components. 
 
We have also proposed linearly converging extensions of IAAL for problems with convex inequality constraints. These are based on a nonquadratic augmented Lagrangian approach such as the exponential, and its dual version, which is an incremental aggregated entropy algorithm \entropyiap. The fuller investigation of this method, as well as the method \entropyiag, which is the exponential analog of the IAG method for nonnegativity constraints, are important subjects for investigation.

\vskip-1.5pc

\section{References}
\vskip-0.9pc
\def\ref{\vskip1.pt\pn}

\ninepoint

\ref[AFB06] Ahn, S., Fessler, J., Blatt, D., and Hero, A.\ O., 2006.\ ``Convergent Incremental Optimization Transfer Algorithms: Application to Tomography," IEEE Transactions on Medical Imaging, Vol.\ 25, pp.\ 283-296.

\ref[BLY15] Bragin, M.\ A., Luh, P.\ B., Yan, J.\ H., Yu, N., and Stern, G.\ A., 2015.\ ``Convergence of the Surrogate Lagrangian Relaxation Method," J.\ of Optimization Theory and Applications, Vol.\ 164, pp.\  173-201.

\ref[BNO03] Bertsekas, D.\ P., Nedi\'c, A., and Ozdaglar, A.\ E., 2003.\
Convex Analysis and Optimization,  Athena Scientific, Belmont, MA.

\ref[BPC11] Boyd, S., Parikh, N., Chu, E., Peleato, B., and Eckstein, J., 2011.\ 
Distributed Optimization and Statistical Learning via the Alternating Direction Method of Multipliers, Now Publishers Inc, Boston, MA.

\ref
[BeT89] Bertsekas, D.\ P., and Tsitsiklis, J.\ N., 1989.\ Parallel and
Distributed Computation: Numerical Methods, Prentice-Hall, Englewood Cliffs,
N.\ J.

\ref
[BeT03] Beck, A., and Teboulle, M., 2003.\ ``Mirror Descent and Nonlinear Projected Subgradient Methods for Convex Optimization," Operations Research Letters, Vol.\ 31, pp.\ 167-175.

\ref[Ber78] Bertsekas, D.\ P., 1978.\ ``Local Convex Conjugacy and Fenchel Duality," Preprints of
7th Triennial World Congress of IFAC, Helsinki, Finland, Vol. 2, pp.\ 1079-1084.

\ref[Ber79] Bertsekas, D.\ P., 1979.\ ``Convexification Procedures and Decomposition Methods for Nonconvex Optimization Problems," J.\ of Optimization Theory and Applications, Vol.\ 29, pp.\ 169-197.

\ref [Ber82] Bertsekas, D.\ P., 1982.\  Constrained Optimization and Lagrange Multiplier Methods, Academic Press, NY; republished in 1996 by Athena Scientific, Belmont, MA. 
 On line at http://web.mit.edu/dimitrib/www/\-lagrmult.html.
 
\ref[Ber09] Bertsekas, D.\ P.,  2009.\ Convex Optimization Theory, Athena
Scientific, Belmont, MA.

\ref[Ber10] Bertsekas, D.\ P., 2010.\ ``Incremental Gradient, Subgradient, and Proximal Methods for Convex Optimization: A Survey," Lab.\ for Information and Decision Systems Report LIDS-P-2848, MIT; arXiv:1507.01030.

\ref[Ber11] Bertsekas, D.\ P., 2011.\ ``Incremental Proximal Methods for Large Scale Convex Optimization," Math.\ Programming, Vol.\ 129, pp.\ 163-195.

\ref[Ber15] Bertsekas, D.\ P.,  2015.\ Convex Optimization Algorithms, Athena
Scientific, Belmont, MA.

\ref [CHH14] Chen, C., He, B., Ye, Y., and Yuan, X., 2014.\ ``The Direct Extension of ADMM for Multi-Block Convex Minimization Problems is not Necessarily Convergent," Mathematical Programming, published on line.

\ref[ChT94] Chen, G., and Teboulle, M., 1994.\ ``A Proximal-Based Decomposition Method for Convex Minimization Problems," Mathematical Programming, Vol.\ 64, pp.\ 81-101.

\ref[DLP14] Deng, W., Lai, M.\ J., Peng, Z., and Yin, W., 2014.\ ``Parallel Multi-Block ADMM with O (1/k) Convergence," arXiv preprint arXiv:1312.3040v2.

\ref[DaL15] Dang, C., and Lan, G., (2015).\ ``Randomized First-order Methods for Saddle Point Optimization," arXiv preprint arXiv:1409.8625v3.

\ref[EcB92] Eckstein, J.,  and Bertsekas, D.\ P., 1992.\ ``On the Douglas-Rachford Splitting Method and the Proximal Point Algorithm for Maximal Monotone Operators," Math.\
Programming, Vol.\ 55, pp.\ 293-318.

\ref[Eck12] Eckstein, J., 2012.\ ``Augmented Lagrangian and Alternating Direction Methods for Convex Optimization: A Tutorial and Some Illustrative Computational Results," RUTCOR Research Report RRR 32-2012, Rutgers, Univ.

\ref[Eve63] Everett, H., 1963.\  ``Generalized Lagrange Multiplier Method for
Solving Problems of Optimal Allocation of Resources," Operations Research, Vol.\ 11, pp.\
399-417.

\ref[FAJ14] Feyzmahdavian, H.\ R., Aytekin, A., and Johansson, M., 2014.\ ``A Delayed Proximal Gradient Method with Linear Convergence Rate," in Prop.\ of 2014 IEEE International Workshop on Machine Learning for Signal Processing (MLSP), pp.\ 1-6.

\ref[GOP15] Gurbuzbalaban, M., Ozdaglar, A., and Parrilo, P., 2015.\ ``On the Convergence Rate of Incremental Aggregated Gradient Algorithms," arXiv preprint arXiv:1506.02081.

\ref[GaM76] Gabay, D., and Mercier, B., 1976.\ ``A Dual Algorithm for the Solution of Nonlinear Variational Problems via Finite-Element Approximations," Comp.\ Math.\ Appl., Vol.\ 2, pp.\ 17-40.

\ref[Gab79] Gabay, D., 1979.\ Methodes Numeriques pour l'Optimization Non Lineaire, These de Doctorat d'Etat et Sciences Mathematiques, Uni.\ Pierre at Marie Curie (Paris VI).

\ref[Gab83] Gabay, D., 1983.\ ``Applications of the Method of Multipliers to Variational Inequalities," in M.\ Fortin and R.\ Glowinski, eds., Augmented Lagrangian Methods: Applications to the Solution of Boundary-Value Problems, North-Holland, Amsterdam.

\ref[GiB15] Giselsson, P., and Boyd, S., 2015.\ ``Metric Selection in Douglas-Rachford Splitting and ADMM," arXiv preprint arXiv:1410.8479v4.

\ref[GlM75]  Glowinski, R. and Marrocco, A., 1975.\ ``Sur l' Approximation par
Elements Finis d' Ordre un et la Resolution par Penalisation-Dualite
d'une Classe de Problemes de Dirichlet Non Lineaires"  Revue
Francaise d'Automatique Informatique Recherche Operationnelle,
Analyse Numerique, R-2, pp.\ 41-76.

\ref[HCW14] Hong, M., Chang, T.-H., Wang, X., Razaviyayn, M., Ma, S., and Luo, Z.-Q., 2013.\ ``A Block Successive Upper Bound Minimization
Method of Multipliers for Linearly Constrained Convex Optimization,? arXiv preprint arXiv:1401.7079v1.

\ref[HaM11] Hamdi, A., and Mishra, S.\ K., 2011.\ ``Decomposition Methods Based on Augmented Lagrangians: A Survey," in Topics in Nonconvex Optimization,  Springer, N.\ Y., pp.\ 175-203.

\ref[HoL13] Hong, M., and Luo, Z.\ Q., 2013.\ ``On the Linear Convergence of the Alternating Direction Method of Multipliers," arXiv preprint arXiv:1208.3922v3.

\ref[IST94] Iusem, A.\ N., Svaiter, B.\ F., and Teboulle, M., 1994.\ ``Entropy-Like Proximal
Methods in Convex Programming," Math.\ of Operations Research, Vol.\ 19, pp.\ 790-814.

\ref[Ius99] Iusem, A.\ N., 1999.\ ``Augmented Lagrangian Methods and Proximal Point 
Methods for Convex Minimization,"
Investigacion Operativa, Vol.\ 8, pp.\ 11-49.

\ref[JuN11a] Juditsky, A., and Nemirovski, A., 2011.\ ``First Order Methods for Nonsmooth Convex Large-Scale Optimization, I: General Purpose Methods," in Optimization for Machine Learning, by Sra, S., Nowozin, S., and Wright, S.\ J.\ (eds.),  MIT Press, Cambridge, MA, pp.\  121-148.

\ref[JuN11b] Juditsky, A., and Nemirovski, A., 2011.\ ``First Order Methods for Nonsmooth Convex Large-Scale Optimization, II: Utilizing Problem's Structure," in Optimization for Machine Learning, by Sra, S., Nowozin, S., and Wright, S.\ J.\ (eds.),  MIT Press, Cambridge, MA, pp.\  149-183.

\ref [KoB72] Kort, B.\ W., and Bertsekas, D.\ P., 1972.\  ``A New Penalty
Function Method for Constrained Minimization," Proc.\ 1972 IEEE Confer.\ Decision Control, New Orleans,
LA, pp.\ 162-166.

\ref[LiM79] Lions, P.\ L., and Mercier, B., 1979.\ ``Splitting Algorithms for the Sum of Two Nonlinear Operators," SIAM J.\ on Numerical Analysis, Vol.\ 16, pp.\ 964-979.

\old{
\ref[LiW14] Liu, J., and Wright, S.\ J., 2014.\ ``Asynchronous Stochastic Coordinate Descent: Parallelism and Convergence Properties," Univ.\ of Wisconsin Report, arXiv preprint arXiv:1403.3862.
}

\ref[Mai13] Mairal, J., 2013.\ ``Optimization with First-Order Surrogate Functions," arXiv preprint arXiv:1305.3120.

\ref[Mai14] Mairal, J., 2014.\ ``Incremental Majorization-Minimization Optimization with Application to Large-Scale Machine Learning," arXiv preprint arXiv:1402.\-4419.

\old{
\ref[MYF03] Moriyama, H., Yamashita N., and Fukushima, M., 2003.\ ``The Incremental Gauss-Newton Algorithm with Adaptive Stepsize Rule," 
Computational Optimization and Applications,
Vol.\ 26, pp.\ 107-141.
}

\ref[Mar70]  Martinet, B., 
1970.\ ``Regularisation d' In\' equations Variationelles par
Approximations Successives," Revue Fran.\ d'Automatique 
et Infomatique Rech.\ Op\' erationelle, Vol.\ 4, pp.\ 154-159.

\ref[NBB01] Nedi\'c, A., Bertsekas, D.\ P., and Borkar, V., 2001.\ ``Distributed Asynchronous Incremental Subgradient Methods," Proc.\ of 2000 Haifa Wor\-kshop ``Inherently Parallel Algorithms in Feasibility and Optimization and Their Applications," by D.\ Butnariu, Y.\ Censor, and S.\ Reich, Eds., Elsevier, Amsterdam.

\ref[NeB01]
Nedi\'c, A., and Bertsekas, D.\ P., 2001.\
``Incremental Subgradient Methods for Nondifferentiable Optimization,''
SIAM J.\ on Optimization,
Vol.\ 12, 2001, pp.\ 109-138.

\ref[NeB10] Nedi\'c, A., and Bertsekas, D.\ P., 2010.\
``The Effect of Deterministic Noise in Subgradient Methods," Math.\ Programming, Ser.\  A, Vol.\ 125, pp.\ 75-99. 

\ref[Ned11] Nedi\'c, A., 2011.\ ``Random Algorithms for Convex Minimization Problems," Math.\  Programming, Ser. B, Vol.\ 129, pp.\ 225-253.

\ref[Nes04] Nesterov, Y., 2004.\  Introductory Lectures on Convex Optimization, Klu\-wer Academic Publisher,
Dordrecht, The Netherlands.

\ref[RoT15] Robinson, D.\ P., and Tappenden, R.\ E., 2015.\ ``A Flexible ADMM Algorithm for Big Data Applications," arXiv preprint arXiv:1502.04391.

\ref [Roc73] Rockafellar, R.\ T., 1973.\  ``A Dual Approach to Solving
Nonlinear Programming Problems by Unconstrained Optimization," Math.\
Programming, pp.\ 354-373.

\ref [Roc76a] Rockafellar, R.\ T., 1976.\  ``Monotone Operators and the
Proximal Point Algorithm," SIAM J.\ on Control and Optimization, Vol.\ 14, pp.\
877-898.

\ref [Roc76b] Rockafellar, R.\ T., 1976.\ ``Augmented Lagrangians and
Applications of the Proximal Point Algorithm in Convex Programming," Math.\ of Operations Research,
Vol.\ 1, pp.\ 97-116.

\ref [Rus95]  Ruszczynski, A., 1995.\ ``On Convergence of an Augmented Lagrangian
Decomposition Method for Sparse Convex Optimization," Math.\ of Operations Research,
Vol.\ 20, pp.\ 634-656.

\ref[SLB13] Schmidt, M., Le Roux, N., and Bach, F., 2013.\ ``Minimizing Finite Sums with the Stochastic
Average Gradient," arXiv preprint arXiv:1309.2388.

\ref[TaM85] Tanikawa, A., and Mukai, M., 1985.\ ``A New Technique for Nonconvex Primal-Dual Decomposition of a Large-Scale Separable Optimization Problem," IEEE Trans.\ Autom.\ Control, Vol.\ AC-30, 
pp.\ 133-143

\ref[Tad89] Tatjewski, P., 1989.\ ``New Dual-Type Decomposition Algorithm for Nonconvex Separable Optimization Problems," Automatica, Vol.\ 25, pp.\  233-242.

\ref [TsB93]  Tseng, P., and Bertsekas, D.\ P., 1993.\ ``On the Convergence of the Exponential
Multiplier Method for Convex Programming," Math.\ Programming, Vol.\ 60, pp.\ 1-19.

\ref[WHM13] Wang, X., Hong, M.,  Ma, S., Luo, Z.\ Q., 2013.\ ``Solving Multiple-Block Separable Convex Minimization Problems Using Two-Block Alternating Direction Method of Multipliers," arXiv preprint arXiv:1308.5294.

\ref[WaB13] Wang, M.,  and Bertsekas, D.\ P., 2013.\ ``Incremental Constraint Proje\-ction-Proximal Methods for Nonsmooth Convex Optimization," Lab.\ for Information and Decision Systems Report LIDS-P-2907, MIT, to appear in SIAM J.\ on Optimization.

\ref[WaB15] Wang, M., and Bertsekas, D.\ P., 2015.\ ``Incremental Constraint Projection Methods for Variational Inequalities," Mathematical Programming, Vol.\  150, pp.\ 321-363.

\vfill\eject
\end
\pn ON WHY THE EXPONENTIAL ITERATION HAS LINEAR CONVERGENCE RATE

\pn As illustration, consider the one-dimensional version of the aggregated incremental iteration
$$x^j_{k+1}=x_k^j \exp\lf(-\a^j \sum_{i=1}^m {\partial f_{i}(x_{\ell_i})\over \partial x^j}\ri),\qquad j=1,\ldots,n,$$
cf.\ Eq.\ \entropyiag. It takes the form
$$x_{k+1}=x_k\exp\lf(-\a \sum_{i=1}^m \gr f_{i}(x_{\ell_i})\ri).\eqno(1)$$
Its nonaggregated gradient version is
$$x_{k+1}=x_k\exp\big(-\a \gr f(x_k)\big).\eqno(2)$$
where $f=f_1+\cdots+f_m$.

Assume that $f$ is strongly convex and has its unique minimum over $x\ge0$ at $x^*$, and that if $x^*=0$ then $\gr f(x^*)>0$. We claim that if $\a$ is constant but sufficiently small and the nonincremental iteration (2) converges to $x^*$, then it converges linearly. The proof admits an extension to prove that if the aggregated incremental version (1) converges to $x^*$, then it converges linearly, as stated in p.\ 32 of my paper.

\proof (Informal; I don't know of a published proof, but the line of proof is based on a standard convergence proof argument for the unconstrained gradient method, with a separate treatment of the components of $x$ that converge to 0.) Consider two cases:
\nitem{(1)} $x^*=0$. Then from Eq.\ (2), we have asymptotically
$$x_{k+1}\approx \g x_k,$$
where
$$\g=\exp\big(-\a \gr f(x^*)\big).$$
For sufficiently small $\a$, we have $\g<1$ [since $\gr f(x^*)>0$], so the convergence of $x_k$ to $x^*$ is linear.

\nitem{(2)} $x^*>0$. Then the first order Taylor expansion of the function
$$p(x)=\exp\big(-\a \gr f(x)\big)$$
around $x=x^*$ yields
$$p(x)= p(x^*)+\gr p(x^*)(x-x^*)+O\big(|x-x^*|^2\big).$$
Since $\gr f(x^*)=0$, we have $p(x^*)=1$ and
$$\gr p(x^*)=-\a \gr^2 f(x^*)\exp\big(-\a\gr f(x^*)\big)=-\a \gr^2 f(x^*),$$
so 
$$\exp\big(-\a \gr f(x_k)\big)=p(x_k)=1-\a \gr^2 f(x^*)(x_k-x^*)+ O\big(|x_k-x^*|^2\big).$$
Using this expression, the iteration (2) can be written as
$$x_{k+1}= x_k\big(1-\a\gr^2 f(x^*)(x_k-x^*)\big)+x_k O\big(|x_k-x^*|^2\big).$$
Equivalently
$$x_{k+1}-x^*= \big(1-\a x_k\gr^2 f(x^*)\big)(x_k-x^*)+x_k O\big(|x-x^*|^2\big).$$
This shows that asymptotically we have
$$|x_{k+1}-x^*|\approx \big|1-\a x^*\gr^2 f(x^*)\big||x_k-x^*|,$$
and since $x^*>0$ and $\gr^2 f(x^*)>0$ (by strong convexity), the convergence is linear for $\a$ sufficiently small so that $\big|1-\a x^*\gr^2 f(x^*)\big|<1$.

As an exercise, it is worth writing up this proof for the case where $x$ is multidimensional, and also for the case of an aggregated incremental method like (1) rather than the nonincremental method (2). This will involve proper accounting of the errors due to incrementalism and errors due to the components of $x_k$ that converge to 0. Of course you should also try to understand the errors in your arguments.

\end

%% file: TEXSHOP_macros_new.tex
%%%%%%%%%%%% MACROS FOR PAPERS AND REPORTS %%%%%%%%%%%%
%My little macros

\def\ignore#1{}
 
%%%%%%%%%%%%%%%%%%%%%

\newcount\sectnum
\newcount\subsectnum
\newcount\eqnumber

\global\eqnumber=1\sectnum=0

% Equation labels

\def\lab{(\the\sectnum.\the\eqnumber)}

%Example of use: suppose we want to give a label \lgh to an equation
% $$ ......  \xdef\lgh{\lab} \eqnum \show{lgh}$$
% Later refer to Eq. \lgh\ ...
% Note the \ after \lgh; it seems to be needed if we want the equation number
% to be followed by a space; not needed if followed by . or ,

%The next macro is used to display labels in drafts, so that you do
%not have to remember them

\def\show#1{#1}

%The next macro is to be used for final drafts that do not display labels
%\def\show#1{}

%%%%%%%%%%%%%%%%%%%%%

\def\smskip{\vskip 5 pt}
\def\medskip{\vskip 10 pt}
\def\bigskip{\vskip 15 pt}
\def\pn{\par\noindent}
\def\br{\break}

\def\bl{\bigl} 
\def\br{\bigr} 
\def\lf{\left}
\def\ri{\right}

\def\frac#1#2{{#1\over #2}}

\def\ol#1{\overline{#1}}

\def\tr{ ^{\prime}}

\def\a{\alpha}

\def\b{\beta}
\def\l{\lambda}
\def\g{\gamma}
\def\m{\mu}

\def\r{\rho}
\def\e{\epsilon}

\def\d{\delta}
\def\s{\sigma}

\def\re{\Re}
\def\rn{\Re^n}

\def\gr{\nabla}

 %break line; horizontal space
\def\tl{\tilde}

\def\old#1{}% invalidates text in braces 
\def\leaderfill{\leaders\hbox to 1em{\hss.\hss}\hfill}
%Example of use: \line{1. Optimality Conditions\leaderfill p.\ 2}

% John's macros

\parindent=2pc
\baselineskip=15pt
\vsize=8.7 true in
\voffset=0.125 true in
\parskip=3pt

% vector/matrix macros

%eqalign macros
\def\minprob#1#2#3{$$\eqalign{&\hbox{minimize\ \ }#1\cr &\hbox{subject to\ \
}#2\cr}\ifnum 0=#3{}\else\eqno(#3)\fi$$}        
     
\def\maxprob#1#2#3{$$\eqalign{&\hbox{maximize\ \ }#1\cr &\hbox{subject to\ \
}#2\cr}\ifnum 0=#3{}\else\eqno(#3)\fi$$}        
     
\def\aligntwo#1#2#3#4#5{$$\eqalign{#1&#2\cr #3&#4\cr}
\ifnum 0=#5{}\else\eqno(#5)\fi$$}
\def\alignthree#1#2#3#4#5#6#7{$$\eqalign{#1&#2\cr #3&#4\cr #5&#6\cr}
\ifnum 0=#7{}\else\eqno(#7)\fi$$}

% Macros to automatically advance equation and other numbers

\def\eqnum{\eqno{\hbox{(\the\sectnum.\the\eqnumber)}\global\advance\eqnumber
by1}}

\def\eqnu{\eqno{\hbox{(\the\sectnum.\the\eqnumber)}\global\advance\eqnumber
by1}}

\newcount\examplnumber
\def\examplnum{\global\advance\examplnumber by1}

\newcount\figrnumber
\def\figrnum{\global\advance\figrnumber by1}

\newcount\propnumber
\def\propnum{\global\advance\propnumber by1}

\newcount\defnumber
\def\defnum{\global\advance\defnumber by1}

\newcount\lemmanumber
\def\lemmanum{\global\advance\lemmanumber by1}

\newcount\assumptionnumber
\def\assumptionnum{\global\advance\assumptionnumber by1}

\newcount\conditionnumber
\def\conditionnum{\global\advance\conditionnumber by1}

\def\exampl{\the\sectnum.\the\examplnumber}
\def\figr{\the\sectnum.\the\figrnumber}
\def\propn{\the\sectnum.\the\propnumber}
\def\defn{\the\sectnum.\the\defnumber}
\def\lemman{\the\sectnum.\the\lemmanumber}
\def\assumptionn{\the\sectnum.\the\assumptionnumber}
\def\condn{\the\sectnum.\the\conditionnumber}

\def\section#1{\goodbreak\vskip 3pc plus 6pt minus 3pt\leftskip=-2pc
   \global\advance\sectnum by 1\eqnumber=1
\global\examplnumber=1\figrnumber=1\propnumber=1\defnumber=1\lemmanumber=1\assumptionnumber=1 \conditionnumber =1%
   \line{\hfuzz=1pc{\hbox to 3pc{\bf %\the\sectnum.\quad
   \vtop{\hfuzz=1pc\hsize=38pc\hyphenpenalty=10000\noindent\uppercase{\the\sectnum.\quad #1}}\hss}}
			\hfill}
			\leftskip=0pc\nobreak\tenf
			\vskip 1pc plus 4pt minus 2pt\noindent\ignorespaces}

% ETP Macros

%\def\section#1{\goodbreak\vskip 3pc plus 6pt minus 3pt\leftskip=-2pc
%   \global\advance\sectnum by 1\eqnumber=1
%   \line{\hfuzz=1pc{\hbox to 3pc{\bf %\the\sectnum.\quad
%   \vtop{\hfuzz=1pc\hsize=38pc\hyphenpenalty=10000\noindent\uppercase{#1}}\hss}}
%                        \hfill}
%                        \leftskip=0pc\nobreak\tenf
%                        \vskip 1pc plus 4pt minus 2pt\noindent\ignorespaces}

\def\sect#1{\noindent\leftskip=-2pc\tenf
   \goodbreak\vskip 1pc plus 4pt minus 2pt
                \global\advance\subsectnum by 1\eqnumber=1
   \line{\hfuzz=1pc{\hbox to 3pc{\bf %\the\sectnum.\quad
   \vtop{\hfuzz=1pc\hsize=38pc\hyphenpenalty=10000\noindent\uppercase{{\bf #1}}}\hss}}
                        \hfill}
   \leftskip=0pc\nobreak\tenf
                        \vskip 1pc plus 4pt minus 2pt\nobreak\noindent\ignorespaces}

\def\subsection#1{\noindent\leftskip=0pc\tenf
   \goodbreak\vskip 1pc plus 4pt minus 2pt
%               \global\advance\subsectnum by 1
   \line{\hfuzz=1pc{\hbox to 3pc{\bf %\the\sectnum.\quad
   \vtop{\hfuzz=1pc\hsize=38pc\hyphenpenalty=10000\noindent{\bf #1}}\hss}}
                        \hfill}
   \leftskip=0pc\nobreak\tenf
                        \vskip 1pc plus 4pt minus 2pt\nobreak\noindent\ignorespaces}
\def\subsubsection#1{\goodbreak\vskip 1pc plus 4pt minus 2pt
   \hfuzz=3pc\leftskip=0pc\noindent\tenit #1 \nobreak\tenf\vskip 6pt plus 1pt
                                minus 1pt\nobreak\ignorespaces\leftskip=0pc}
%
%\def\rthl{Sec. \the\chapnum.\the\sectnum}                      
%\def\rthc{#1}\nobreak\noindent\ignorespaces
%\newcount\sectnum \sectnum=0
%\newcount\subsectnum \subsectnum=0

\def\beginexample#1{\noindent\goodbreak\vskip 6pt plus 1pt minus 1pt
\noindent
  \hbox {\bf Example #1\hss}%\break%\noindent
  \nobreak\vskip 4pt plus 1pt minus 1pt \nobreak\noindent\ninef
  \global\advance
                \leftskip by\parindent\pn}
\def\endexample{\vskip 12pt\tenf\par
  \global\advance\leftskip by -\parindent
  }

\def\beginexercise#1{\noindent\goodbreak\vskip 6pt plus 1pt minus 1pt \noindent\global\normalbaselineskip=12pt
  \hbox {\bf Exercise #1\hss}%\break%\noindent
  \nobreak\vskip 4pt plus 1pt minus 1pt 
  \nobreak\noindent\ninef\global\advance\leftskip
                        by\parindent\pn}
\def\endexercise{\vskip 12pt\tenf\par
  \global\advance\leftskip by -\parindent
  }

\def\beginsection#1{\noindent\goodbreak\vskip 6pt plus 1pt minus 1pt \noindent\global\normalbaselineskip=12pt
  \hbox {\it #1\hss}
  \vskip 0.1pt plus 1pt minus 1pt \nobreak\noindent\ninef\global\advance
                \leftskip by\parindent\noindent\pn}
\def\endsection{\vskip 12pt\tenf\par
  \global\advance\leftskip by -\parindent
}

%

% Header/Title macros

\def\lemma#1{\smskip\pn{\bf Lemma #1}\quad}

\def\proposition#1{\smskip\pn{\bf Proposition #1}\quad}
\def\proof{\smskip\pn{\bf Proof:}\quad}

 \def\qed{\quad{\bf
Q.E.D.} \par\bigskip}
\def\ref{\smskip\pn}

\def\chapter#1#2{{\bf \centerline{\helbigbig
{#1}}}\bigskip\bigskip{\bf \centerline{\helbigbig
{#2}}}\bigskip\bigskip} % ex. \chapter{Chapter 1}{Title of chapter}

 % ex. \longchapter{Chapter 1}{Title of chapter}{Title of
 %chapter}

 % ex. \papertitle{Title of paper}{Names of Authors}

\def\longpapertitle#1#2#3{{\bf \centerline{\helbigb
{#1}}}\bigskip{\bf \centerline{\helbigb
{#2}}}\bigskip\bigskip{\centerline{
by}}\bigskip{\bf \centerline{
{#3}}}\bigskip\bigskip} 
% ex. \longpapertitle{First part of title of paper}
%{2nd part of title of paper}{Names of Authors}

% List macros

\def\nitem#1{\smskip\item{#1}}

\newcount\alphanum
\newcount\romnum

\def\alphaenumerate{\ifcase\alphanum \or (a)\or (b)\or (c)\or (d)\or (e)\or
(f)\or (g)\or (h)\or (i)\or (j)\or (k)\fi}
\def\romenumerate{\ifcase\romnum \or (i)\or (ii)\or (iii)\or (iv)\or (v)\or
(vi)\or (vii)\or (viii)\or (ix)\or (x)\or (xi)\fi}

\def\alist{\begingroup\vskip10pt\alphanum=1% alphabetical list
\parskip=2pt\parindent=0pt \leftskip=3pc
\everypar{\llap{{\rm\alphaenumerate\hskip1em}}\advance\alphanum by1}}

\def\nolist{\begingroup\vskip10pt\alphanum=0% numerical list
\parskip=2pt\parindent=0pt \leftskip=3pc
\everypar{\llap{\global\advance\alphanum by1(\the\alphanum)\hskip1em}}}

\def\romlist{\begingroup\vskip10pt\romnum=1% roman list
\parskip=2pt\parindent=0pt \leftskip=5pc
\everypar{\llap{{\rm\romenumerate\hskip1em}}\advance\romnum by1}}

% romlist indents more than alist or nolist and can be used inside them

%Figure, table, and box macros

\long\def\fig#1#2#3{\vbox{\vskip1pc\vskip#1
\prevdepth=12pt \baselineskip=12pt
\vskip1pc
\hbox to\hsize{\hfill\vtop{\hsize=25pc\noindent{\eightbf Figure #2\ }
{\eightpoint#3}}\hfill}}}%Figure space definition. Example of use:
%\fig{16pc}{1.1}{A network with one central processor and a separate
%communication link to each device.}

\long\def\widefig#1#2#3{\vbox{\vskip1pc\vskip#1
\prevdepth=12pt \baselineskip=12pt
\vskip1pc
\hbox to\hsize{\hfill\vtop{\hsize=28pc\noindent{\eightbf Figure #2\ }
{\eightpoint#3}}\hfill}}}

\long\def\table#1#2{\vbox{\vskip0.5pc
\prevdepth=12pt \baselineskip=12pt
\hbox to\hsize{\hfill\vtop{\hsize=25pc\noindent{\eightbf Table #1\ }
{\eightpoint#2}}\hfill}}}

%Running Head Macros
 
\def\rightheadline#1{\headline{\tenrm\hfil #1}}

% Concept Macros

\long\def\leftfig#1#2{\vbox{\smskip\hsize=220pt
\vtop{{\noindent {\bf #1}}}
\smskip
\noindent
\vbox{{\noindent #2}}
}}

\long\def\rightfig#1#2#3{\vbox{\smskip\vskip#1
\prevdepth=12pt \baselineskip=12pt
\hsize=210pt
\smskip
\vbox{\noindent{\eightbold #2}
\hskip1em{\eightpoint#3}}
}}

\long\def\concept#1#2#3#4#5{\bigskip\hrule
\vbox{\hbox{\leftfig{#1}{#2} \hskip3em
\rightfig{#3}{#4}{#5}} \smskip}
\hrule\bigskip}

% Example of Use: \concept{Title of Concept}{Text}
% {Figure size}{Figure number?}{Figure caption}

\long\def\bconcept#1#2#3#4#5#6#7{
\vbox{
\hbox to \hsize{\vtop{\par #1}}
\concept{#2}{#3}{#4}{#5}{#6}
\hbox to \hsize{\vtop{\par #7}}
\smskip}
}

% Example of Use: \bconcept{Preceding text}{Title of Concept}{Text}
% {Figure size}{Figure number}{Figure caption}{Following text}

% same as concept but without the \hrule's; ready to be boxed

% Put inside a box

\def\boxit#1{\vbox{\hrule\hbox{\vrule\kern3pt
                                \vbox{\kern3pt#1\kern3pt}\kern3pt\vrule}\hrule}}
% example of use: \setbox0=\vbox{.... }; \boxit{\box0}
\def\centerboxit#1{$$\vbox{\hrule\hbox{\vrule\kern3pt
                                \vbox{\kern3pt#1\kern3pt}\kern3pt\vrule}\hrule}$$}
% example of use: \setbox0=\vbox{.... }; \centerboxit{\box0}

\long\def\boxtext#1#2{$$\boxit{\vbox{\hsize #1\noindent\strut #2\strut}}$$}
% example of use: \boxtext{462pt}{This is the boxed text.}; 462pt is max length

% Picture macros and examples
%
% figures must be pasted from mcdraw
%
% Look in the 'Windows' menu for the pictures window
% It's like the Scrapbook -- cut and paste pictures
%

\def\picture #1 by #2 (#3){
  \vbox to #2{
    \hrule width #1 height 0pt depth 0pt
    \vfill
    \special{picture #3} % this is the low-level interface
    }
  }
% The first dimension of the picture macro is the width the second is depth

\def\scaledpicture #1 by #2 (#3 scaled #4){{
  \dimen0=#1 \dimen1=#2
  \divide\dimen0 by 1000 \multiply\dimen0 by #4
  \divide\dimen1 by 1000 \multiply\dimen1 by #4
  \picture \dimen0 by \dimen1 (#3 scaled #4)}
  }

%
% Note that you can also say, e.g.,
%  \special{postscript xxx yyy zzz}
% to include PostScript graphics in your documents
%
%Examples of use
%\def\stripes{\picture 2.29in by 1.75in (AWstripes)}
% By executing \stripes 
%\def\annie{\scaledpicture 102pt by 239pt (annie scaled 2000)}
%\def\finder{\picture 260pt by 165pt (screen0 scaled 500)}
%\def\icon{\picture 7in by 7in (icon)}
%Example of use
%\annie
%Example of centered picture \line{\hfil\annie\hfil}

%Figure w/  caption macro
\long\def\captfig#1#2#3#4#5{\vbox{\vskip1pc
\hbox to\hsize{\hfill{\picture #1 by #2 (#3)}\hfill}
\prevdepth=9pt \baselineskip=9pt
\vskip1pc
\hbox to\hsize{\hfill\vtop{\hsize=24pc\noindent{\eightbold Figure #4}
\hskip1em{\eightpoint#5}}\hfill}}}

%Examples of use of Figure macros
%\captfig{8.53pc}{19.9pc}{picturename}{5}{caption.}
%\captfig{2.23in}{2in}{picturename scaled 500}{16}{Caption.}
%The macro centers the picture.
%The first two numbers should be the true width
% and height after the picture has been scaled.
% So if the picture is scaled by 50% (500), the width and height in
% the macro should onw half of what they would be if the picture
% is not scaled (1000).
%
%
%
% Postcript macros

\def\illustration #1 by #2 (#3){
  \vskip#2\hskip#1\special{illustration #3} % this is the low-level interface
    }

\def\scaledillustration #1 by #2 (#3 scaled #4){{
  \dimen0=#1 \dimen1=#2
  \divide\dimen0 by 1000 \multiply\dimen0 by #4
  \divide\dimen1 by 1000 \multiply\dimen1 by #4
  \illustration \dimen0 by \dimen1 (#3 scaled #4)}
  }

% SHADEBOX.BSR MACROS
% Author: Leo@vaxc.cc.monash.edu.au
% Original Source:  Posted by Jimm Herreron <HERRERON@SMCVAX.BITNET>
% Modified from the file SHADEBOX.TEX on 9/30/93 by Becky Kaluza of Blue Sky
% Research to work with Textures 1.5 or later.

\newbox\graybox
\newdimen\xgrayspace
\newdimen\ygrayspace
%
% This macro can be used to typeset some text in a framed box with a
% shaded background. A set of examples can be found at the end of this
% file.
%
% This is a plain \TeX\ file modified for use on the Macintosh with Textures
% 1.5 or later.
%
% The characteristics of the shaded boxes are controlled by the following
% parameters
%
%   \xgrayspace = the space added before and after the text
%   \ygrayspace = the space above and below the text
%   \grayshade  = the gray colour 0 = black 1 = white
%   \linewidth  = the thickness of the border in points
%   \theradius  = the radius of the rounded corners in points
%   \thevskip   = extra \vskip added above and below the shaded box
%                 (applies only to \parashade)
%
%----------------------------------------------------------------------------
%
% The following \TeX code was adapted from previous work by
%
%            Je'ro^me Maillot, maillot@bora.inria.fr
%----------------------------------------------------------------------------
%
% Use the following for one or more words within a line.
%

\def\Textshade#1#2#3#4#5#6{%
    \xgrayspace=#4pt%
    \ygrayspace=#4pt%
    \def\grayshade{#3}%
    \def\linewidth{#5}%
    \def\theradius{#6}%
    \setbox\graybox=\hbox{\surroundboxa{#2}}%
    \hbox{%
    \hbox to 0pt{%
%!    \special{"gsave newpath 0 0 moveto                                %
    \PScommands
}%
    \box\graybox}}%
%
% Use the following for paragraphs.
%
\long%

\long%
\def\Parashade#1#2#3#4#5#6#7{%
    \xgrayspace=#4pt%
    \ygrayspace=#4pt%
    \def\grayshade{#3}%
    \def\linewidth{#5}%
    \def\theradius{#6}%
    \def\thevskip{#7pt}%
    \setbox\graybox=\hbox{\surroundboxb{#2}}%
    \vskip\thevskip%
    \hbox{%
    \hbox to 0pt{%
%!    \special{"gsave newpath 0 0 moveto                                %
    \PScommands
}%
     \box\graybox}%
     \vskip\thevskip%
}%
%----------------------------------------------------------------------------
%
% A pair of box macros. Each builds a slightly oversized box
% containing the text. The text is centred both in the vertical
% horizontal directions.
%
% Use the following for one or more words within a line.
%
\long\def\surroundboxa#1{\leavevmode\hbox{\vtop{%
\vbox{\kern\ygrayspace%
\hbox{\kern\xgrayspace#1%
      \kern\xgrayspace}}\kern\ygrayspace}}}
%
% Use the following for a paragraphs.
%
\long\def\surroundboxb#1{\leavevmode\hbox{\vtop{%
\vbox{\kern\ygrayspace%
\hbox{\kern\xgrayspace\vbox{\advance\hsize-2\xgrayspace#1}%
      \kern\xgrayspace}}\kern\ygrayspace}}}
%----------------------------------------------------------------------------
%
% Here are some simple PostScript routines.
%
% The TeX command \PScommands must be called before any of the
% shading routines can be used.
%
%!\long\def\PScommands{\special{! TeXDict begin
\long\def\PScommands{%
\special{rawpostscript
/sharpbox{%
           newpath
           xmin ymin moveto
           xmin ymax lineto
           xmax ymax lineto
           xmax ymin lineto
           xmin ymin lineto
           closepath 
          }bind def
}%
\special{rawpostscript
/sharpboxnb{%
           newpath
           xmin ymin moveto
           xmin ymax lineto
           xmax ymax lineto
           xmax ymin lineto
%           xmin ymin lineto
%           closepath 
          }bind def
}%
\special{rawpostscript
/sharpboxnt{%
           newpath
           xmin ymax moveto
           xmin ymin lineto
           xmax ymin lineto
           xmax ymax lineto
%           xmin ymin lineto
%           closepath 
          }bind def
}%
\special{rawpostscript
/roundbox{%
           newpath
           xmin radius add ymin moveto
           xmax ymin xmax ymax radius arcto
           xmax ymax xmin ymax radius arcto
           xmin ymax xmin ymin radius arcto
           xmin ymin xmax ymin radius arcto 16 {pop} repeat
           closepath
          }bind def
}%
\special{rawpostscript
/sharpcorners{%
               sharpbox gsave grayshade setgray fill grestore 
               linewidth setlinewidth stroke
              }bind def
}%
\special{rawpostscript
/sharpcornersnt{%
               sharpboxnt gsave grayshade setgray fill grestore 
               linewidth setlinewidth stroke
              }bind def
}%
\special{rawpostscript
/sharpcornersnb{%
               sharpboxnb gsave grayshade setgray fill grestore 
               linewidth setlinewidth stroke
              }bind def
}%
\special{rawpostscript
/roundcorners{%
               roundbox gsave grayshade setgray fill grestore 
               linewidth setlinewidth stroke
              }bind def
}%
\special{rawpostscript
/plainbox{%
           sharpbox grayshade setgray fill 
          }bind def
}%
% Here are the two new options
%
\special{rawpostscript
/roundnoframe{%
               roundbox grayshade setgray fill 
              }bind def
}%
\special{rawpostscript
/sharpnoframe{%
               sharpbox grayshade setgray fill 
              }bind def
}%
%!end}%
}%
%
% The \PScommands macro must be invoked before the shaded box macros.
%
%!\PScommands
% To use this, type \textshade{plainbox} or \textshade{roundbox} or
% \textshade{sharpbox}

\def\pshade#1{\Parashade{sharpcorners}{#1}{0.95}{10}{0.5}{10}{10}}

%%%%% BOXES FOR TEXSHOP %%%%%

\def\boxit#1{\vbox{\hrule\hbox{\vrule\kern3pt
                                \vbox{\kern3pt#1\kern3pt}\kern3pt\vrule}\hrule}}
% example of use: \setbox0=\vbox{.... }; \boxit{\box0}

\def\boxitnb#1{\vbox{\hrule\hbox{\vrule\kern3pt
                                \vbox{\kern3pt#1\kern3pt}\kern3pt\vrule}}}

\def\boxitnt#1{\vbox{\hbox{\vrule\kern3pt
                                \vbox{\kern3pt#1\kern3pt}\kern3pt\vrule}\hrule}}

\long\def\boxtext#1#2{$$\boxit{\vbox{\hsize #1\noindent\strut #2\strut}}$$}
% example of use: \boxtext{462pt}{This is the boxed text.}; 462pt is max length

% example of use: \boxtext{462pt}{This is the boxed text.}; 462pt is max length

% example of use: \boxtext{462pt}{This is the boxed text.}; 462pt is max length

\def\texshopbox#1{\boxtext{462pt}{\vskip-1.5pc\pshade{\vskip-1.0pc#1\vskip-2.0pc}}}

%***************************************************
%         FONTS
%***************************************************

% ROMAN
\font\hel=cmr10%
\font\helbigbig=cmr10 scaled 2500%
\font\helbigb=cmbx10 scaled 1500%
\font\eightbold=cmbx8%

\def\tenf{\hel}%
\def\tenit{\heli}%
\def\ninef{\ninehel}%
\def\nineit{\nineheli}%
%
%

%  FONT FAMILIES

\font\tenrm=cmr10%
\font\teni=cmmi10%
\font\tensy=cmsy10%
\font\tenbf=cmbx10%
\font\tentt=cmtt10%
\font\tenit=cmti10%
\font\tensl=cmsl10%

\def\tenpoint{\def\rm{\fam0\tenrm}%
\textfont0=\tenrm%
\textfont1=\teni%
\textfont2=\tensy%
\textfont\itfam=\tenit%
\textfont\slfam=\tensl%
\textfont\ttfam=\tentt%
\textfont\bffam=\tenbf%
\scriptfont0=\sevenrm%
\scriptfont1=\seveni%
\scriptfont2=\sevensy%
%\scriptfont3=\tenex%
\scriptscriptfont0=\sixrm%
\scriptscriptfont1=\sixi%
\scriptscriptfont2=\sixsy%
%\scriptscriptfont3=\tenex%
\def\it{\fam\itfam\tenit}%
\def\tt{\fam\ttfam\tentt}%
\def\sl{\fam\slfam\tensl}%
\scriptfont\bffam=\sevenbf%
\scriptscriptfont\bffam=\sixbf%
\def\bf{\fam\bffam\tenbf}%
\normalbaselineskip=18pt%
\normalbaselines\rm}%

\font\ninerm=cmr9%
\font\ninebf=cmbx9%
\font\nineit=cmti9%
\font\ninesy=cmsy9%
\font\ninei=cmmi9%
\font\ninett=cmtt9%
\font\ninesl=cmsl9%

\def\ninepoint{\def\rm{\fam0\ninerm}%
\textfont0=\ninerm%
\textfont1=\ninei%
\textfont2=\ninesy%
\textfont\itfam=\nineit%
\textfont\slfam=\ninesl%
\textfont\ttfam=\ninett%
\textfont\bffam=\ninebf%
\scriptfont0=\sixrm%
\scriptfont1=\sixi%
\scriptfont2=\sixsy%
%\scriptfont3=\tenex%
\def\it{\fam\itfam\nineit}%
\def\tt{\fam\ttfam\ninett}%
\def\sl{\fam\slfam\ninesl}%
\scriptfont\bffam=\sixbf%
\scriptscriptfont\bffam=\fivebf%
\def\bf{\fam\bffam\ninebf}%
\normalbaselineskip=16pt%
\normalbaselines\rm}%

\font\eightrm=cmr8%
\font\eighti=cmmi8%
\font\eightsy=cmsy8%
\font\eightbf=cmbx8%
\font\eighttt=cmtt8%
\font\eightit=cmti8%
\font\eightsl=cmsl8%

\def\eightpoint{\def\rm{\fam0\eightrm}%
\textfont0=\eightrm%
\textfont1=\eighti%
\textfont2=\eightsy%
\textfont\itfam=\eightit%
\textfont\slfam=\eightsl%
\textfont\ttfam=\eighttt%
\textfont\bffam=\eightbf%
\scriptfont0=\sixrm%
\scriptfont1=\sixi%
\scriptfont2=\sixsy%
%\scriptfont3=\tenex%
\scriptscriptfont0=\fiverm%
\scriptscriptfont1=\fivei%
\scriptscriptfont2=\fivesy%
%\scriptscriptfont3=\tenex%
\def\it{\fam\itfam\eightit}%
\def\tt{\fam\ttfam\eighttt}%
\def\sl{\fam\slfam\eightsl}%
%\scriptfont\bffam=\sixbf%
\scriptscriptfont\bffam=\fivebf%
\def\bf{\fam\bffam\eightbf}%
\normalbaselineskip=14pt%
\normalbaselines\rm}%

\font\sevenrm=cmr7%
\font\seveni=cmmi7%
\font\sevensy=cmsy7%
\font\sevenbf=cmbx7%

\font\sixrm=cmr6%
\font\sixi=cmmi6%
\font\sixsy=cmsy6%
\font\sixbf=cmbx6%

\fontdimen13\tensy=2.6pt%
\fontdimen14\tensy=2.6pt%
\fontdimen15\tensy=2.6pt%
\fontdimen16\tensy=1.2pt%
\fontdimen17\tensy=1.2pt%
\fontdimen18\tensy=1.2pt%       

\def\tenf{\tenpoint}%
\def\ninef{\ninepoint}%
%

%%%%%%%%%%%% END OF MACROS %%%%%%%%%%%%